\documentclass[10pt,aps,prx,twocolumn,superscriptaddress,floats,floatfix,showpacs,longbibliography]{revtex4-1}

\usepackage[T1]{fontenc}
\usepackage{graphicx}
\usepackage{color}
\usepackage{upgreek}
\usepackage{setspace}
\usepackage{amssymb}
\usepackage{amsmath}
\usepackage{pstricks}
\usepackage{dsfont}
\usepackage{fancyhdr}
\usepackage{array}
\usepackage{multirow}
\usepackage{booktabs}
\usepackage{diagbox}
\usepackage{hhline}
\usepackage{footmisc}
\usepackage{float}
\usepackage{soul}
\usepackage{hyperref}

\usepackage[normalem]{ulem} 
\usepackage{amsfonts}
\usepackage{txfonts}
\usepackage{lipsum}
\usepackage{wasysym}
\usepackage{etoolbox} 
\usepackage{bbold}
\usepackage{mathtools} 
\usepackage{multirow} 
\usepackage{hhline}
\usepackage{mathrsfs} 

\DeclareMathOperator{\Tr}{Tr}

\newcommand{\av}[1]{\ensuremath{\left\langle #1 \right\rangle}}
\newcommand{\qv}{\mathbf{q}}
\newcommand{\kv}{\mathbf{k}}

\usepackage{letltxmacro}
\LetLtxMacro{\oldsqrt}{\sqrt}
\renewcommand{\sqrt}[2][\mkern8mu]{\mkern-6mu\mathop{}\oldsqrt[#1]{#2}}

\allowdisplaybreaks[4]

\newcolumntype{M}[1]{>{\centering\arraybackslash}m{#1}}

\global\long\def\av#1{\left\langle #1 \right\rangle }

\global\long\def\cd{c^\dag}

\global\long\def\c{{\bm c}}

\global\long\def\eps{\varepsilon}
\global\long\def\part{\partial}
\global\long\def\Tr{{\rm Tr}}

\begin{document}
\title{Collective magnetic fluctuations in Hubbard plaquettes captured by fluctuating local field method}

\author{Alexey N. Rubtsov}
\email{ar@rqc.ru}
\affiliation{Russian Quantum Center, Skolkovo innovation city, 121205 Moscow, Russia}
\affiliation{Department of Physics, Lomonosov Moscow State University, Leninskie gory 1, 119991 Moscow, Russia}

\author{Evgeny  A.  Stepanov}
\affiliation{I. Institute of Theoretical Physics, Department of Physics, University of Hamburg, Jungiusstrasse 9, 20355 Hamburg, Germany}
\affiliation{Theoretical Physics and Applied Mathematics Department, Ural Federal University, Mira Str. 19, 620002 Ekaterinburg, Russia}

\author{Alexander I. Lichtenstein}
\affiliation{I. Institute of Theoretical Physics, Department of Physics, University of Hamburg, Jungiusstrasse 9, 20355 Hamburg, Germany}
\affiliation{European XFEL, Holzkoppel 4, 22869 Schenefeld, Germany}
\affiliation{Theoretical Physics and Applied Mathematics Department, Ural Federal University, Mira Str. 19, 620002 Ekaterinburg, Russia}

\begin{abstract}
    We establish a way to handle main collective fluctuations in correlated quantum systems based on a Fluctuation Local Field concept.
    This technique goes beyond standard mean-field approaches, such as Hartree-Fock and dynamical mean-field theories (DMFT), as it includes a fluctuating classical field that acts on the leading order parameter of the system.
    Effective model parameters of this new theory are determined from the variational principle, which allows to resolve the Fierz ambiguity in decoupling of the local interaction term.
    In the saddle-point approximation for the fluctuating field our method reproduces the mean-field result. 
    The exact numerical integration over this field allows to consider nonlinear fluctuations of the global order parameter of the system while local correlations can be accounted by solving the DMFT impurity problem.
    We apply our method to the magnetic susceptibility of finite Hubbard systems at half-filling and demonstrate that the introduced technique leads to a superior improvement of results with respect to parental mean-field approaches without significant numerical complications. 
    We show that the Fluctuation Local Field method can be used in a very broad range of temperatures substantially below the N\'eel temperature of DMFT, which remains a major challenge for all existing theoretical approaches.
\end{abstract}

\maketitle

\section{Introduction}

The theoretical description of collective effects of interacting fermionic systems is one of the main problems of modern physics. 
In correlated materials, these collective electronic fluctuations form effective bosonic modes, such as plasmons, magnons, and etc, that may possess a nonlinear behavior. 
The origin of the latter can be both, the interaction between different modes, as well as the anharmonic fluctuation of the single mode itself. 
At low temperatures, the presence of these instability channels may result in a spontaneous symmetry breaking associated with the formation of ordered phases in the system. 
Strong collective fluctuations appear not only in infinite crystal lattices, but also in other physical systems that are not necessarily large and can be essentially finite.
In this regard, one can mention vibrational modes in nuclei~\cite{kamal2014nuclear}, breathing modes~\cite{abraham2014quantum} and short-range charge and spin correlations~\cite{Greif1307, hart2015observation, PhysRevLett.115.260401, Cheuk1260} in systems of ultracold atoms trapped in optical lattices, and collective spin modes in molecular magnets~\cite{blanc2018quantum, coronado2019molecular, holynska2019single}. 

A large collection of theoretical approaches from simple mean-field theories~\cite{Fradkin, STB, You_1982, Holden_1982} and rotationally invariant path-integral schemes~\cite{PhysRevB.19.2626, PhysRevB.20.4584, PhysRevLett.61.467, PhysRevLett.65.2462, PhysRevB.43.3790, ScheurerE3665, PhysRevB.52.R11557, Shulz_PI}, to much more advanced methods~\cite{RevModPhys.90.025003} has been developed in order to describe these collective effects. The simplest mean-field theory, namely the Hartree-Fock (HF) method~\cite{PhysRev.35.210.2, Fock1930}, is able to capture a spontaneous symmetry breaking in weakly interacting systems. For strong correlations, the preference is often given to a polarized dynamical mean-field theory (DMFT)~\cite{RevModPhys.68.13}. 
This approach relies on the exact numerical solution of an effective impurity problem, which provides an accurate approximation for local observers~\cite{PhysRevB.91.235114}. 
Various diagrammatic extensions of DMFT have been constructed to handle nonlocal correlations underlying the formation of collective electron modes~\cite{RevModPhys.90.025003}.  
A particular subset of diagrams varies for different methods, and may contain either simple $GW$-like diagrams~\cite{PhysRevB.66.085120, PhysRevLett.90.086402, PhysRevLett.109.226401, PhysRevB.87.125149, PhysRevB.90.195114, PhysRevB.94.201106, PhysRevB.95.245130} including vertex corrections~\cite{PhysRevB.92.115109, PhysRevB.93.235124, PhysRevLett.119.166401, PhysRevB.100.205115}, or more complex ladder~\cite{PhysRevB.77.033101, PhysRevB.75.045118, PhysRevB.80.075104, Rubtsov20121320, PhysRevB.90.235135, PhysRevB.93.045107, PhysRevB.94.205110, PhysRevB.100.165128} and parquet~\cite{PhysRevB.101.075109} contributions, as well as all possible diagrammatic terms up to a certain order of perturbation expansion~\cite{PhysRevB.94.035102, PhysRevB.96.035152, 2020arXiv200704669V}.

Diagrammatic techniques introduced on top of DMFT are currently seen as the most advanced and promising tools for description of collective fluctuations in correlated systems. It should be mentioned however, that applicability of these computational scheme is limited, because it implicitly exploits an assumption of a weak anharmonicity of collective modes. 
Indeed, collective nonlocal effects in these approaches are considered perturbatively, so that only (nearly) harmonic fluctuations of the corresponding order parameter are taken into account. 
This assumption works reasonably well at high temperatures and/or in high dimensions~\cite{PhysRevLett.113.246407, PhysRevB.90.235105, PhysRevB.94.205110, NatureVanLoon, NatureZhenya}. 
However, strong collective fluctuations that suppose to break the ordering at low temperatures are strongly nonlinear. 
It can be expected, that such physics is particularly important for antiferromagnetic (AFM) fluctuations, since the AFM order parameter obeys strong quantum zero-point fluctuations even for 3D bulk materials, whereas in 2D the ordering at finite temperatures is forbidden by Mermin-Wagner theorem~\cite{PhysRevLett.17.1133}. 
We also believe that the nonlinearity of collective fluctuations is a fundamental reason why all the introduced above diagrammatic extensions of DMFT, as well as exact diagrammatic Monte Carlo approaches~\cite{PhysRevLett.81.2514, Kozik_2010, PhysRevLett.119.045701, PhysRevB.93.161102, PhysRevB.100.121102, sigmaDet}, do not allow for a quantitative description of the 2D Hubbard model substantially below the DMFT N\'eel point~\cite{2020arXiv200610769S}. 
In this regard, the desired improvement of the theory for strong collective fluctuations especially at low temperatures calls for new ideas.

In this work we turn off the beaten path of doing a perturbative expansion around the mean-field result, and present a completely new method that is capable of description of non-linear collective fluctuations in finite systems at low temperatures. 
For the sake of concreteness, we apply the formulated theory to small Hubbard plaquettes. 
Among all mentioned systems that possess strong collective fluctuations, the considered model is mostly close to molecular magnets. 
However, we would like to point out that the present scheme is rather general and can be constructed for any system where developed collective fluctuations belong to the one or several leading collective modes.
In addition, the result for small Hubbard lattices at half-filling can also be efficiently benchmarked, because
Quantum Monte Carlo (QMC) simulations for such lattices do not suffer from the sign problem~\cite{PhysRevB.41.9301} and therefore can provide a reliable reference data. 

We note that the correct description of the AFM susceptibility of these small systems, which we aim to address in this paper, is already a challenging problem. 
For example, our calculations show that for the $4\times 4$ square plaquette with periodic boundary conditions applicability of the best local mean-field approximation, namely DMFT, is limited to a temperature of about $U/8$, where $U$ is the on-site Coulomb potential. 
At a twice smaller temperature, this approach that neglects nonlocal correlations shows an unphysical N\'eel transition. 
At the same time, a spatial pattern of AFM fluctuations on finite plaquettes is expected to be rather simple. 
In this case, the small number of lattice sites results in a coarse grid 
for the Brillouin zone. Then, already a single AFM mode, which is associated with the $Q = \{\pi, \pi\}$ momentum, should capture the most important physics of spin fluctuations in the system.
Fluctuations with other wave vectors in small systems are not important, because the AFM correlation length is larger than the systems size. 

\begin{figure}[t!]
\includegraphics[width=0.90\linewidth]{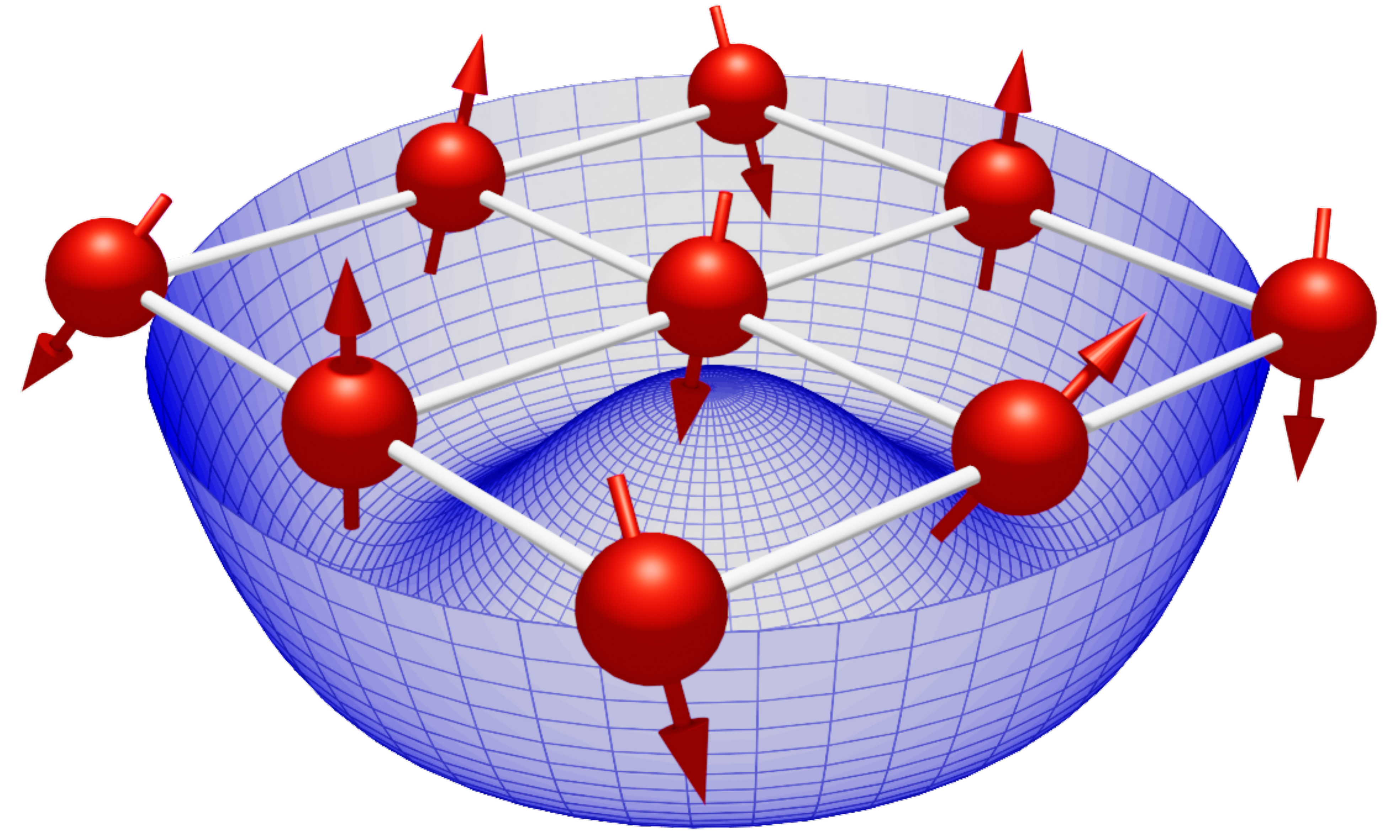} 
\caption{\label{Sketch} Sketch of spin-fluctuations (red arrows) in a Hubbard plaquette near the antiferromgnetic state, which is described by an effective Mexican-hat potential (blue). }
\end{figure}

Consequently, one can introduce a Landau free energy ${\cal F}(m)$, where magnetization of the AFM sublattices $m$ serves as a ``global'' order parameter. 
According to Landau phenomenology, at low temperatures the second derivative of the free energy $\partial^2_{m} {\cal F}(m)|_{m=0}$ becomes negative, and ${\cal F}(m)$ takes the form of a Mexican-hat potential (see Fig.~\ref{Sketch}).  
As has been mentioned above, diagrammatic schemes built on top of DMFT partly take collective electronic effects into account introducing a renormalization for a corresponding two-particle fluctuation.
However, consideration of a leading subset 
of diagrams implies that a small nonlinearity of fluctuations is assumed. 
Taking strong nonlinearity into account formally requires to sum over all diagrams, as it is done, for example, in diagrammatic QMC calculations~\cite{PhysRevLett.81.2514, Kozik_2010, PhysRevLett.119.045701, PhysRevB.93.161102, PhysRevB.100.121102, sigmaDet}. 
Note is that this exact perturbation expansion does not converge at low temperatures, and a renormalization procedure is required to achieve physically interesting regimes~\cite{PhysRevB.96.041105}.
Thus we conclude, that diagrammatic schemes on the basis of DMFT are justified until a Mexican-hat potential is formed~\cite{2020arXiv200610769S}. 
At the same time, their applicability at lower temperatures, where fluctuations of the order parameter become essentially unharmonic, is questionable. In addition, we find diagrammatic extensions of DMFT technically too complicated for a rather simple system under investigation. 
Instead, here we propose a solution of the problem introducing an effective local field that mediates nonlinear fluctuations of the order parameter we are interested in.
To this aim we build a Fluctuating Local Field (FLF) approach on the basis of two mean-field schemes starting from Hartree-Fock and DMFT solutions of the problem.
Previously, a similar approach has been invented for classical lattices~\cite{PhysRevE.97.052120}.  
The introduced FLF method is numerically inexpensive and does not bring a sufficient complication to its parental mean-field theory. In particular, the FLF scheme built on top of DMFT does not involve any calculation of two-particle vertices of the impurity problem, or similar quantities.  
At the same time, nonlinear collective AFM fluctuations are explicitly included in this scheme and accounted exactly. Thus, the theory unperturbatively accounts for both, local correlations and collective fluctuations by solving the DMFT impurity problem and integrating over the fluctuating field.
We compare our results to the QMC reference data for $4\times 4$, $6 \times 6$, and $8\times8$ square plaquettes, and demonstrate that the introduced FLF technique leads to an impressive improvement of mean-field results.

\section{Theory}

\subsection {Definitions}

Our goal is to address spin fluctuations of small correlated lattices. Whereas the following consideration can be applied to quite wide class of systems, we stick to a particular case of a half-filled Hubbard model on a square lattice with the following action:
\begin{align}
\label{Hubbard}
{\cal S}[\cd,c] &= - \cd_{1} {\cal G}^{-1}_{12} c^{\phantom{*}}_{2}  + U 
\left(n_{j\tau\uparrow}-\frac{1}{2}\right)\left(n_{j\tau\downarrow}-\frac{1}{2}\right).  
\end{align}
Here, $c^{(\dagger)}$ are Grassmann variables corresponding to the annihilation (creation) of electrons. Subscripts ``1'' and ``2'' are combined indices of the lattice site $j$ (or momentum ${\bf k}$), imaginary time $\tau$ (or Matsubara frequency $\omega=\pi(2p+1)/\beta$, $p\in\mathbb{Z}$), and spin projection $\sigma=\{\uparrow, \downarrow\}$.
$n_{j\tau\sigma}=c^{\dagger}_{j\tau\sigma} c^{\phantom{*}}_{j\tau\sigma}$ describes the electron density, and $\beta$ is the inverse temperature. 
Through the paper, the tensor notation is used, so that the summation is taken over repeated indices. 
The bare Green's function reads 
\begin{equation}
\label{eq:Gbare}
{\cal G}^{-1} = i \omega - \eps + h^l \Lambda^l_{Q}, 
\end{equation}
where the dispersion $\eps_{\bf k}= - 2 t (\cos k_x + \cos k_y)$ is the Fourier transform of the nearest-neighbor hopping amplitude $t$. 
Note that here, we additionally account for an external AFM field ${\bf h}$. 
The tensor $\Lambda_{Q}$ describes the coupling of the field $ \bf h$ to fermion degrees of freedom. Its explicit structure  can be seen from the following expression 
\begin{align}
\cd_1 {\cal G}^{-1}_{12} c^{\phantom{*}}_2 = \cd_{{\bf k}\omega\sigma} \left(i \omega - \eps^{\phantom{*}}_{\bf k}\right) c^{\phantom{*}}_{{\bf k}\omega\sigma} + \cd_{{\bf k}\omega\sigma} \sigma_{\sigma\sigma'}^l h^l c^{\phantom{*}}_{{\bf k}+Q,\omega,\sigma'},
\end{align}
where $Q = \{\pi, \pi\}$ is the AFM wave vector, and $\sigma^{l}$ is the ${l=\{x, y, z\}}$ component of the vector of Pauli matrices.
The field $\bf h$ is then coupled to the AFM-ordered variable ${M^{l}_{Q} =  \cd_1 \Lambda^{l}_{12} c^{\phantom{*}}_2}$ that describes polarization of electrons. Then, the average ${\bf m} = \frac{1}{\beta{}N}\av{{\bf M}}$ is the AFM order parameter of the system. 
While constructing the formalism, it is convenient to keep $\bf h$ finite. A paramagnetic solution can be obtained taking the limit ${\bf h} \to 0$ afterwards.   

\subsection{Resolution of Fierz ambiguity}
\label{sec:Fierz}

The goal of the FLF approach is to identify the leading instability in the system, which is to be accounted exactly neglecting other less important modes. The former procedure has to be performed carefully, because it may lead to hidden problems such as Fierz ambiguity~\cite{PhysRevD.68.025020, PhysRevB.70.125111, Jaeckel}. To illustrate this point, let us consider the initial problem~\eqref{Hubbard} in a mean-field approximation. For this aim we rewrite the on-site Coulomb potential in terms of spin densities as
\begin{align}
\label{Hubbard_ss}
U \left(n_{j\tau\uparrow}-\frac{1}{2}\right)\left(n_{j\tau\downarrow}-\frac{1}{2}\right) = -\frac{1}{2}  s^l_{j \tau} {\cal U}_{ll'} s^{l'}_{j\tau}, 
\end{align}
where 
$s^l = \cd_{\sigma}  \sigma^l_{\sigma\sigma'} c^{\phantom{*}}_{\sigma'}$ is the $l$ projection of the spin density, and ${\cal U}$ is an arbitrary symmetric $3\times 3$ matrix with the constrained trace $\Tr\,{\cal U} = U$. 
In what follows we assume for simplicity that ${\cal U}$ has an inverse. 
In special cases when this is not the true, one can consider some approximation or a particular block of ${\cal U}$ that is invertible.
This decoupling of the local Coulomb interaction can also be done with inclusion of charge degrees of freedom. However, the latter are not of the interest for the current work, because they do not represent the main instability of the considered system.  

The mean-field description of spin degrees of freedom can be performed introducing an effective vector field $\varphi$ via Hubbard-Stratonovich transformation of the interaction term~\eqref{Hubbard_ss}. The partition function of the problem can now be rewritten as ${\cal Z} = \int D[\varphi]D[\cd, c] \, e^{-{\cal S}[\cd, \,c, \,\varphi]}$, where
\begin{align}
{\cal S}[\cd, c, \varphi] 
&= - \cd_{1} {\cal G}^{-1}_{12} c_{2} - \varphi_{j\tau}^l s^l_{j\tau} + \frac{1}{2} \varphi^l_{j\tau} {\cal U}^{-1}_{ll'} \varphi^{l'}_{j\tau}.
\label{eq:Sp}
\end{align}
The mean-field value $\varphi_{\rm MF}$ can be obtained from the saddle-point approximation for the integral over the vector field $\varphi$, which can be expressed in the following condition
\begin{equation}
\delta {\cal S}[\varphi_{\rm MF}]=0.
\label{eq:saddle}
\end{equation}
Here, the action for Hubbard-Stratonovich fields  
\begin{equation}\label{eq:Landau-Fierz}
{\cal S}[\varphi] = -   \ln \det  \left[{\cal G}^{-1} +  \varphi^l \Lambda^l_{Q} \right] + \frac{1}{2}   \varphi^l_{\tau j}  {\cal U}^{-1}_{ll'} \varphi^{l'}_{\tau j}
\end{equation}
can be obtained after integrating out fermion degrees of freedom in Eq.~\eqref{eq:Sp} leading to ${\cal Z} = \int D[\varphi] e^{-{\cal S}[\varphi]}$. This results in $\varphi^{l}_{\rm MF} = \frac12{\cal U}_{ll'}\langle s^{l'}\rangle$, and
we get a mean-field approximation for the action~\eqref{eq:Sp}
\begin{align}
{\cal S}_{\rm MF}[\cd, c] &= - \cd_{1} {\cal G}^{-1}_{12} c_{2} - \varphi_{\rm MF}^l M^l_{Q}
\label{eq:SMF}
\end{align}
that describes non-interacting fermions in the presence of an effective field  
\begin{align}
h^l_{eff}= h^l+\frac12{\cal U}_{ll'} m^{l'}_{\rm MF}.
\label{eq:phisp}
\end{align}
Here, ${\bf m}_{\rm MF}$ is the average AFM magnetization of the mean-field problem~\eqref{eq:SMF}.
Now, it becomes clear that the obtained result crucially depends on a particular choice of the matrix ${\cal U}$ leading to Fierz ambiguity in decoupling of the interaction term. Indeed, considering the spin polarization along the $z$-axis, we get $\varphi^{z}_{sp} = \frac12{\cal U}_{zz}m^{z}$, where ${\cal U}_{zz}$ in the isotropic decoupling form ${\cal U}_{xx}={\cal U}_{yy}={\cal U}_{zz}=U/3$ is three times smaller than in the Ising form, where only $z$ component of the spin is considered ${\cal U}_{zz}=U$, ${\cal U}_{xx}={\cal U}_{yy}=0$. 

Remarkably, not only simple mean-field theories suffer from the Fierz ambiguity. This issue is also present in more elaborate methods like the $GW$+EDMFT approach~\cite{PhysRevB.66.085120,PhysRevLett.90.086402, PhysRevLett.109.226401, PhysRevB.87.125149, PhysRevB.90.195114, PhysRevB.94.201106, PhysRevB.95.245130} and the triply irreducible local expansion (TRILEX)~\cite{PhysRevB.92.115109, PhysRevB.93.235124, PhysRevLett.119.166401} that have been introduced to solve strongly interacting electronic problems.
A physical reason for Fierz ambiguity is that Hubbard-Stratonovich fields $\varphi$ exhibit strong fluctuations, which make the saddle point approximation for the integral inaccurate. 
All theories that treat the interaction in a mean-field form effectively perform an expansion around the saddle-point approximation.
Different decouplings of the on-site Coulomb potential~\eqref{Hubbard_ss} produce different fluctuation patterns and different mean-field solutions for the same initial problem, and it is not {\it a priory} clear which form of the interaction should be chosen. 
In principle, the Fierz ambiguity can be avoided if the interaction is taken in the form that provides the most accurate result for some quantity that can be calculated exactly. 
Recently, this idea has been exploited for the derivation of the interaction for the DMFT-based dual TRILEX (D-TRILEX) method, by approximating the exact renormalized local fermion-fermion interaction~\cite{PhysRevB.100.205115}. Importantly, it was argued that the most accurate form of the effective interaction ${\cal U}_{ll'}$ cannot be obtained by any decoupling of the on-site Coulomb potential.  
Here, we show that the Fierz ambiguity can also be cured by a renormalization of parameters of an effective theory.

Let us consider the problem~\eqref{eq:SMF} as a trial action, where an effective field $\varphi_{eff}$ plays a role of a free parameter that may differ from the saddle-point value $\varphi_{\rm MF}$. This parameter can be chosen, for example, using the Peierls-Feynman-Bogoliubov variational principle~\cite{PhysRev.54.918, bogolyubov1958variation, feynman1972statistical} for the functional 
\begin{align}
\label{eq:variation}
{\cal F}(\varphi_{eff}) = {\cal F}_{\rm MF}(\varphi_{eff}) + (\beta N)^{-1}\av{{\cal S}[\cd, c] - {\cal S}_{\rm MF}[\cd, c]}_{\rm MF}.
\end{align}
Here, ${\cal S}[\cd,c]$ is the initial action~\eqref{Hubbard}, ${\cal F}_{\rm MF}(\varphi_{eff})$ is the free energy of the mean-field action~\eqref{eq:SMF}, and $\av{\ldots}_{\rm MF}$ denotes averaging with respect to the mean-field partition function ${\cal Z}_{\rm MF}$. An optimal value of $\varphi_{eff}$ can be found minimizing the energy $\partial_{\varphi_{eff}}{\cal F}(\varphi_{eff})=0$.
This consideration gives a well-known Hartree-Fock result 
\begin{align}
h^l_{eff}= h^l+\frac12 U m^{l}_{\rm MF}
\label{eq:phieff}
\end{align}
that does not depend on the form of the decoupling~\eqref{Hubbard_ss}.
Importantly, this variational solution of the problem is not only unambiguous, but also known to provide a quantitatively correct result at least for weakly correlated systems in high dimensions.

The above consideration may look rather trivial, but it serves as a very instructive starting point for construction of the fluctuation local field method. 
Indeed, the optimal value of the effective field $\varphi_{eff}$ obtained via Peierls-Feynman-Bogoliubov variational principle does not depend on the decoupling. 
This means, that the same result~\eqref{eq:phieff} can also be obtained in the saddle-point approximation~\eqref{eq:phisp}, but only for one particular decoupling~\eqref{Hubbard_ss}, which in our case corresponds to the Ising form discussed above. 
The saddle-point approximation of an integral is convenient from many points of view.
Therefore, instead of finding a particular decoupling form of the on-site Coulomb potential~\eqref{Hubbard_ss}, we propose to consider 
the following renormalization procedure that improves the saddle-point approximation. 
From a mathematical point of view, this can be performed taking into account a matrix of second derivatives (curvature) of the field $\varphi$ at the $\varphi_{\rm MF}$ point. 
Formally this means that the action that enters the extremum condition~\eqref{eq:saddle} should be changed to ${\cal S}[\varphi] + {\cal S'}[\varphi]$. 
Instead of the explicit calculation of ${\cal S'}[\varphi]$, we assume that it can be accounted by a proper renormalization of the ``stiffness'' ${\cal U}^{-1}$ in Eq.~\eqref{eq:Landau-Fierz} calculated at the saddle point.
Practically, we adjust ${\cal U}$ to get a saddle-point approximation~\eqref{eq:saddle} coinciding with the HF result~\eqref{eq:phieff}. 
For example, for the isotropic decoupling this condition gives ${\cal U}_{xx}={\cal U}_{yy}={\cal U}_{zz}=U$. 
We note, that such a renormalized interaction ${\cal U}_{ll}$ is not trace-constrained anymore and thus cannot be obtained by decoupling the local Coulomb interaction~\eqref{Hubbard_ss}. 
In this sense, the presented idea is consistent with the result of the D-TRILEX method~\cite{PhysRevB.100.205115}.
The use of the renormalized low-energy interaction for Hubbard-Stratonovich fields is one of the key ingredients of the FLF approach presented below.
 
\subsection{FLF on top of Hartree-Fock method}

There are two assumptions underlying the HF theory. First, the interaction $U$ should be small enough to neglect higher-order corrections to an effective field~\eqref{eq:phieff}. Second, even a weakly interacting system can exhibit strong collective fluctuations that are neglected in the HF scheme.
Thus, an improvement of the HF theory would naturally require an account for these collective fluctuations. Following the pathway proposed in Ref.~\onlinecite{PhysRevE.97.052120} this can be done replacing the constant effective HF field $\varphi_{eff}$~\eqref{eq:phieff} by a fluctuating vector field $V$ introducing an {\it ensemble} of effective Gaussian actions  
\begin{align}
{\cal S}_{\rm FLF} [\cd, c, V] 
&= - \cd_{1} {\cal G}^{-1}_{12} c_{2} - V^l M^l_{Q} + \frac12\frac{\beta N}{J_{Q}} {\bf V}^2.
\label{eq:FLF}
\end{align}
Importantly, this action is different from the exact one~\eqref{eq:Sp} that represents the initial theory.   
Here, unlike quantum Hubbard-Stratonovich fields $\varphi_{j\tau}$, we deal with a classical three-component vector field $V$ that describes only the leading magnetic mode with the zero bosonic frequency $\Omega=0$ and AFM momentum $Q=(\pi,\pi)$. Other fluctuations, as well as quantum fluctuations of the isolated AFM mode, are neglected, since the field $V$ does not depend on $j$ and $\tau$.
One can expect, that our approach is particularly relevant for small lattices, where only one discrete ${\bf q}$-mode softens and becomes essentially unharmonic at low temperatures. However, the role of quantum fluctuations described by $\Omega\neq 0$ is {\it a priori} not clear and will be addressed further.
It is important that no assumption is made about the magnitude of AFM fluctuations or whether they are harmonic or not.

We have shown previously, that simple neglection of fluctuations in high-energy modes may lead to incorrect results. Following the receipt obtained in Sec.~\ref{sec:Fierz}, this issue is solved by introducing a renormalization of the interaction via a ``stiffness'' parameter $J_{Q}$.
The former can be chosen in different and, generally speaking, non equivalent ways. First, let us assume for a moment that the integral over $V$ in the partition function 
\begin{align}
{{\cal Z}_{\rm FLF}=\int D[\cd, c] \, d^3 V \, e^{-{\cal S}_{\rm FLF}[\cd, c, V]}}=\int d^3 V \, e^{-{\cal S}_{\rm FLF}[V] }
\label{eq:Z_DMFT}
\end{align}
is estimated from the saddle-point approximation~\eqref{eq:saddle}, where 
\begin{equation}
\label{Fnu}
{\cal S}_{\rm FLF}[V]=-\ln \det \left[{\cal G}^{-1} + V^l \Lambda^l_{Q} \right] + \frac12\frac{\beta N}{J_{Q}} {\bf V}^2.
\end{equation}
Straightforwardly, one gets
\begin{align}
V^l_{\rm MF} = J_{Q} \, m^{l}.
\label{eq:condition}
\end{align}
Physically, the saddle-point approximation means that fluctuations of the field $V$ are neglected. 
It is worth mentioning that the average magnetization $m^{l}$ contains an effect of an external field $h$ via the bare Green's function~\eqref{eq:Gbare}. 
Noting that $V_{\rm MF}$ also acts as a polarized AFM field, it is reasonable to demand that for any value of the external field $h$ the saddle point approximation should reproduce the HF result~\eqref{eq:phieff}, which also does not account for fluctuations of the order parameter.
This immediately results in the $J_{Q}=U/2$ value of a stiffness constant.
Now, when all parameters of the FLF action~\eqref{eq:FLF} are identified, the integral over the field $V$ can be taken numerically exactly after integrating out fermion degrees of freedom.

It is worth noting, that the self-consistent Hartree-Fock result~\eqref{eq:phieff} and, consequently, $V_{\rm MF}$ changes dramatically upon lowering the temperature. 
Whereas at high temperatures an effective field $h_{eff}$ is proportional to the external field $h$, below the HF N\'eel point the average magnetization $m$ is finite even for an infinitesimal $h$. 
Nevertheless, our saddle-point analysis results in the same constant value $J_{Q}=U/2$ within the entire temperature range.

There exists another possibility how the stiffness parameter $J_{Q}$ can be chosen.
Instead of finding the value of $J_{Q}$ that reproduces the Hartree-Fock result~\eqref{eq:phieff}, which is obtained via the Peierls-Feynman-Bogoliubov variational principle~\eqref{eq:variation}, one can directly use $J_{Q}$ as a variational parameter for the mapping of the original model~\eqref{Hubbard} onto a trial action~\eqref{eq:FLF}. 
For this aim, the FLF $V$ in Eq.~\eqref{eq:FLF} can be integrated out directly, which results in the following trial action 
\begin{align}
{\cal S}^{*}_{\rm FLF} [\cd, c] 
&= - \cd_{1} {\cal G}^{-1}_{12} c_{2} - \frac12\frac{J_{Q}}{\beta N} M^l_{Q} M^l_{Q}.
\label{eq:mapping}
\end{align}
Using the same variational principle for the functional~\eqref{eq:variation}, where the mean-field action ${\cal S}_{\rm MF}[\cd, c]$ is now replaced by the trial action ${\cal S}^{*}_{\rm FLF}[\cd, c]$, one analytically gets $J_{Q}\simeq U/2$.
In addition, we also performed a direct numerical minimization of the functional, which gave $J_{Q}\simeq U/2$ as well.
A detailed derivation of this result can be found in Appendix~\ref{app:Variation}.  

To get a further insight into the method and obtain an additional justification of our choice of $J_{Q}$, it is instructive to see how the Landau free energy ${\cal F}(V) = (\beta N)^{-1} {\cal S}(V)$ behaves at $h=0$.
The value of ${\cal F}'(V) = - (\beta N)^{-1} \ln \det \left[ {\cal G}^{-1} + V^l \Lambda^l_{Q} \right]^{-1}$ decreases with an increase of $V$. 
For small $V$, the system responses linearly, so that ${\cal F}'(V\to0) = {\cal F}'(0) - \frac12 \chi_0 V^2$, where $\chi_0$ is the bare static susceptibility of the lattice.
Therefore, ${\cal F}(V)$ exhibits a minimum (maximum) at $V=0$ if $\chi_0$ is smaller (larger) than $J_{Q}^{-1}$. 
It can be shown, that for $J_{Q}=U/2$ the transition between two regimes occurs at the HF N\'eel temperature. 
For large $V$, the spin polarization saturates at some $m_{max}$, so that ${\cal F}'(V\to \infty) = -V \,m_{max}$. 
Therefore, the term $\frac12 J_{Q}^{-1} V^2$ dominates at large $V$, which guarantees a convergence of the integral in ${\cal Z}_{\rm FLF}$.
This qualitative behaviour of ${\cal F}(V)$ resembles what one would expect for the Landau free energy of a phenomenological theory for a critical phenomena. However, there are important differences. 
First, in our consideration ${\cal F}(V)$ is not a function of the order parameter $m$ but of an effective field $V$ that acts on the order parameter. 
Second, ${\cal F}(V)$ is different from the common 2-4 form of the double-well potential. 
In particular, it shows the $\propto V^2$ behaviour at large $V$.

\begin{figure}[t!]
\includegraphics[width=0.95\linewidth]{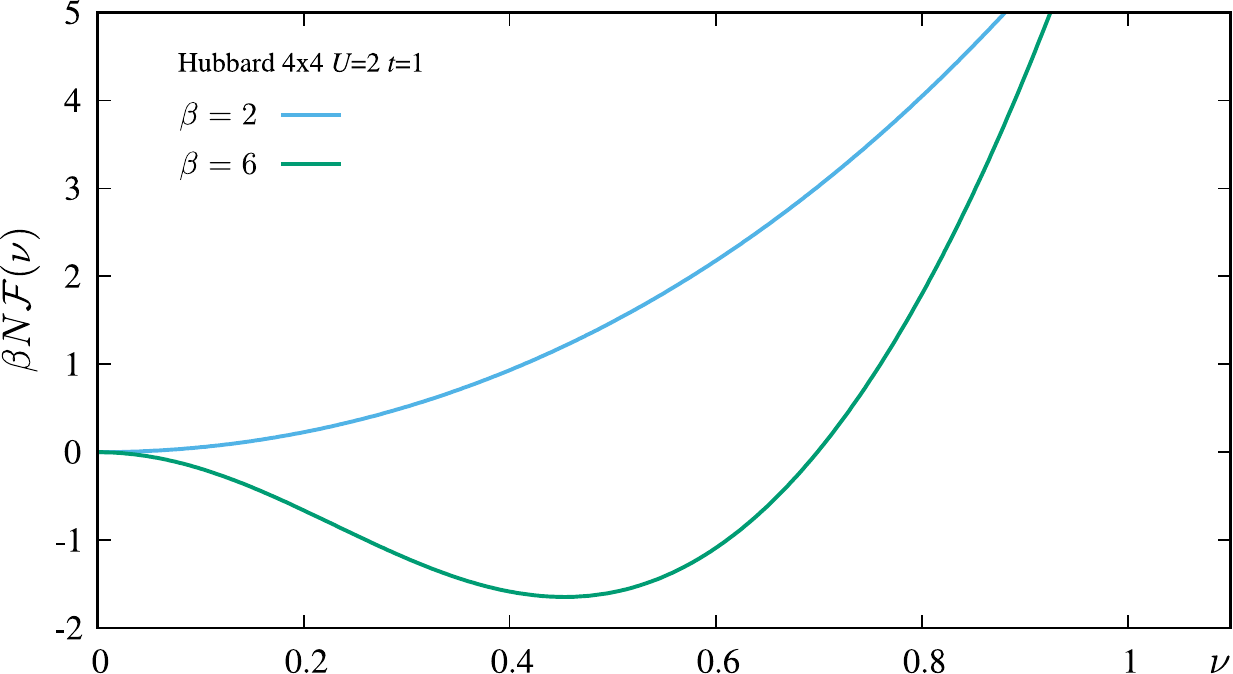}
\caption{\label{figure-HF.Fnu} The value of $\beta N {\cal F}(V)$ for the FLF-HF theory plotted as a function of the field $V$. Upper (blue) and lower (green) curves are obtained for regions below ($\beta=2$) and above ($\beta=6$) the HF N\'eel point $\beta_{\rm N}\approx 3.8$, respectively.}
\end{figure}

Now let us discuss the role of fluctuations using a specific example. Fig.~\ref{figure-HF.Fnu} shows behaviour of the Landau free energy for the $4\times4$ plaquette for $U=2t$. We plot $\beta N {\cal F}(V)$ for two inverse temperatures $\beta=6$ and $\beta=2$ below and above the HF transition point, respectively. 
For high temperature the curve shows a single minimum at $V=0$. 
Below the HF transition point the system reveals phase (Goldstone) fluctuations originated from a degeneracy of the minimum of the Mexican hat potential. 
The amplitude of fluctuations can be estimated from regions where $\beta N {\cal F}(V)$ deviates from its minimal value by $\sim1$. This deviation corresponds to an exponential change of the energy defined by the partition function ${\cal Z}$. 
As one can see, these regions are remarkably broad, although both values $\beta=2$ and $\beta=6$ are not very close to the HF N\'eel point $\beta\approx 3.8$. Thus, it can be concluded that small Hubbard lattices indeed exhibit strong non-Gaussian fluctuations of the order parameter within a broad temperature interval.

\subsection{FLF on top of DMFT approach}

For $U$ larger than several hopping amplitudes $t$, the Hubbard model exhibits strong correlations. 
They are manifested in a local moment formation and appearance of Hubbard subbands in a single-particle spectrum. 
This physics is not captured by Hartree-Fock method. 
In this regime dynamical mean-field theory is more suited to address this problem. 
Within the DMFT, local correlations are taken into account exactly and unperturbatively with the help of an auxiliary local system
\begin{align}
{\cal S}^{(j)}_{\rm imp}[\cd, c] &= -\cd_{j \omega \sigma} (i\omega - \Delta^{\sigma\sigma'}_{j\omega} ) c_{j \omega \sigma'} + U \Big(n_{j\tau\uparrow}-\frac{1}{2}\Big) \Big(n_{j\tau\downarrow}-\frac{1}{2}\Big) 
\label{Simp_new}
\end{align}
Here, the introduced hybridization function $\Delta$ is local in space, which allows to solve this single-site impurity problem exactly and obtain the local Green's function $G_{\rm imp}$. 
The DMFT partition function then reads 
(see Ref.~\onlinecite{RevModPhys.78.865}, and also Appendix~\ref{AppendixB} for different derivation)
\begin{align}
\label{DMFT}
{\cal Z}^{\phantom{1}}_{\rm DMFT} &= {\cal Z}^{\phantom{1}}_{\rm imp} \det G^{\phantom{1}}_{\rm imp} \det G^{-1}_{\rm DMFT} \notag\\ &= {\cal Z}_{\rm imp} \det \left[1+G_{\rm imp} \left(\Delta-\eps+h^l{\Lambda}^l_{Q} \right)\right],    
\end{align}
where ${\cal Z}_{\rm imp}$ is the partition function of the impurity problem~\eqref{Simp_new}, and $G_{\rm DMFT}$ is the DMFT Green's function that can be found from the following relation
\begin{equation}
\label{DMFT-G-new}
G^{-1}_{\rm DMFT}=G^{-1}_{\rm imp} + \Delta - \eps +h^l{\Lambda}^l_{Q}.
\end{equation} 
The hybridization function $\Delta$ is obtained using the self-consistency condition that the local part of the DMFT Green's function is equal to the impurity Green's function.

The important physics that can be captured by DMFT is primarily related with a formation of a local magnetic moment at each lattice site. 
Inclusion of the frequency-dependent hybridization function $\Delta_{\omega}$ allows to account for a formation of the local moment, which is important for the Mott physics.
However, in the paramagnetic regime local moments at different lattice sites are not correlated, and these collective fluctuations are missing at the DMFT level. 
To get an inspiration how the DMFT can be improved, it is instructive to consider a finite Hubbard lattice at low temperatures. 
According to Mermin-Wagner theorem this system is paramagnetic.
Therefore, at $h=0$ one should formally deal with a non-polarized DMFT solution associated with a spin-independent hybridization function $\Delta^{0}$ and Green's function $G^{0}_{\rm imp}$. 
However, this approximation turns to be unsatisfactory, because $G_{imp}^0$ does not contain information about the local magnetic moment of the impurity problem, which exhibits strong fluctuations around its zero average value. 
Moreover, these fluctuations are characterized by a much smaller timescale than a single-particle dynamics described by a single-electron Green's function. 
Thus, it would be physically correct to replace $G_{imp}^0$ in~\eqref{DMFT-G-new} by an {\it ensemble} of polarized Green's functions that provide different realizations of the local spin moment.
For this aim we introduce the FLF-DMFT approach described below. 

Following the strategy we have used to improve the Hartree-Fock theory, we assume that magnetic fluctuations in the system are represented by a classical AFM vector field $V$, so that the fluctuating Green's function of the auxiliary system equals $G_{\rm imp}^0-V^l L^l_{Q}$. Here $L_{Q}$ is a tensor quantity similar to $\Lambda_{Q}$ in Eq.~\eqref{eq:Gbare} that additionally carries an $\omega$ frequency dependence.
Then, the FLF-DMFT partition function can be written as ${{\cal Z}_{\rm FLF}={\cal Z}_{\rm imp} \int d^3 V e^{-{\cal S}_{\rm FLF}[V]}}$ with
\begin{align}
{\cal S}_{\rm FLF}[V] = &-\ln \det \left[1+(G_{\rm imp}^0-V^{l} L^{l}_{Q}) (\Delta^0-\eps+h^{l'}\Lambda^{l'}_{Q})\right] \notag\\
&+ \frac12\frac{\beta N}{J_{Q}}{\bf V}^2 
\label{Fnu2}
\end{align}
where the FLF-DMFT Green's function is (see Appendix~\ref{AppendixB})
\begin{align}
G_{\rm FLF} &= \av{\frac{1}{(G_{\rm imp}^0 - V^{l} L^{l}_{Q})^{-1} + \Delta^0 - \eps + h^{l'} \Lambda^{l'}_{Q}}}_{\rm FLF}.
\label{eq:G_saddle}
\end{align}

\begin{figure*}[ht!]
\includegraphics[width=0.95\linewidth]{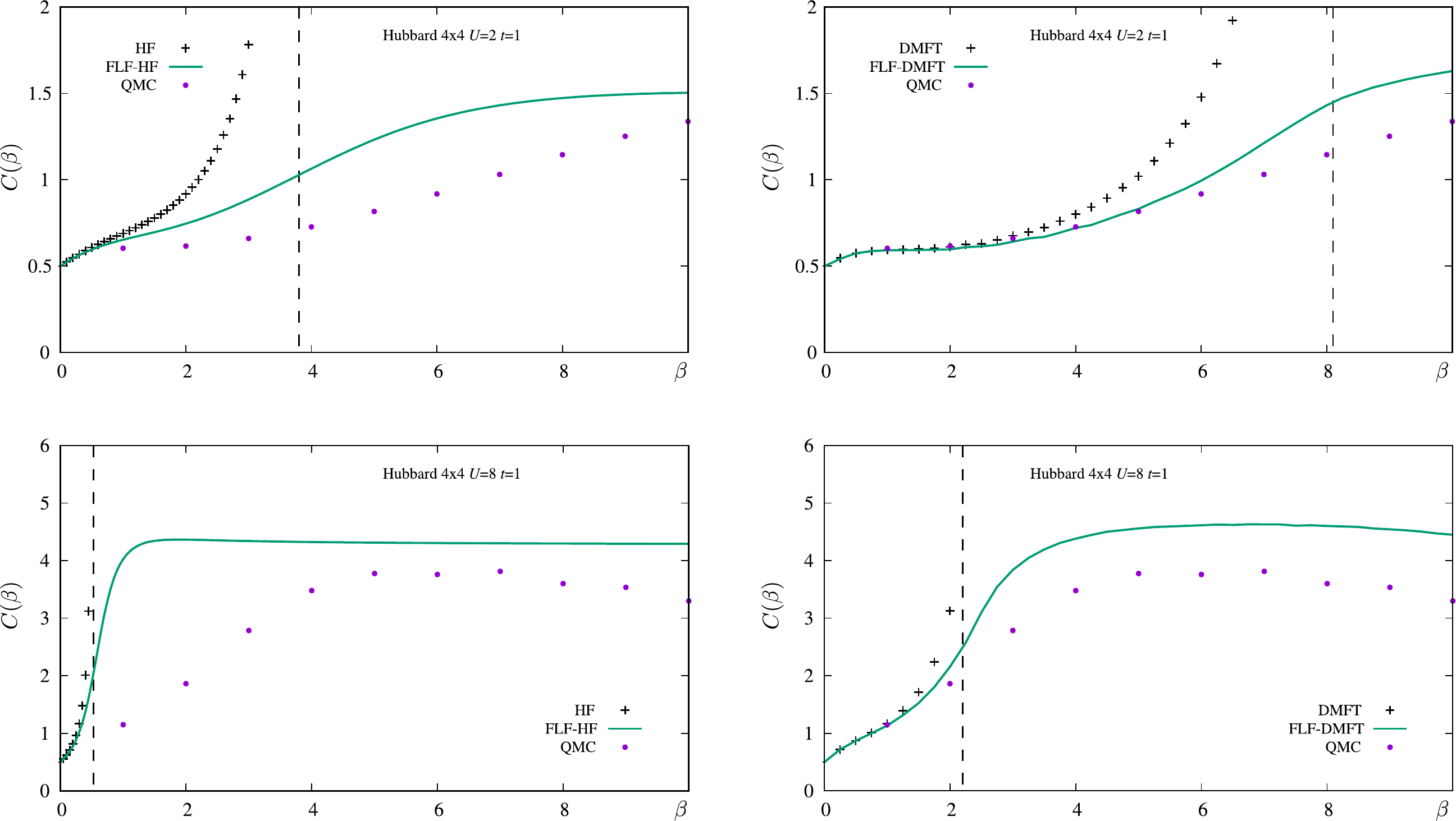} 
\caption{\label{figure-Curie} Curie constant $C(\beta)$ as a function of the inverse temperature $\beta$ calculated for $4\times4$ Hubbard plaquettes with periodic boundary conditions for $U=2t$ (top panels) and $U=8t$ (bottom panels). Results are obtained within HF (left column) and DMFT (right column) schemes. QMC reference data is depicted by purple dots, mean-field (HF and DMFT) results are shown by black pluses, and FLF approach corresponds to a solid green line. Vertical dashed line indicates the mean-field N\'eel transition.}
\end{figure*}

It is important to compare expressions for the FLF-HF~\eqref{Fnu} and FLF-DMFT~\eqref{Fnu2}. 
They are essentially different in the way how the FLF is introduced. 
In FLF-HF fluctuations are associated with an effective field~\eqref{eq:phisp}, which is, in fact, the only adjustable parameter of the HF theory. 
On the contrary, as can be seen from Eq.~\eqref{DMFT-G-new}, there are {\it two} quantities that appear in DMFT -- hybridization function $\Delta$ and Green's function $G_{\rm imp}$. 
One could introduce a theory where fluctuations are associated with the hybridization function. 
In this case, the theory will be similar to the FLF-HF approach, where the saddle-point approximation for the FLF reproduces the polarized local self-energy (see Appendix~\ref{AppendixB}).
However, a physical picture of a fluctuating local moment suggests a different approach expressed by Eq.~\eqref{Fnu2}, where fluctuations of the local Green's function are considered. 
Nevertheless, in the former case a direct connection between the FLF-HF and FLF-DMFT theories can still be established. 
As we show in Appendix~\ref{AppendixB}, this form of the FLF-DMFT approach can be seen as FLF-HF theory introduced for effective fermion variables in a dual space.

The saddle point estimation for the Green's function~\eqref{eq:G_saddle} is
\begin{equation}
\label{DMFT-saddle-new}
G_{\rm FLF}^{\rm MF~-1}= \left(G^{\rm imp}_0 - V_{\rm MF}^l L^l_{Q} \right)^{-1}+\Delta_0-\eps+h^{l'}\Lambda^{l'}_{Q},    
\end{equation}
where $V_{\rm MF}$ can be found from the saddle-point equation
\begin{equation}
\label{lambda-DMFT-new}
J_{Q}^{-1}{\bf V}^2_{\rm MF} = {\rm Tr} \frac{ -V_{\rm MF}^l L^l_{Q}}{(\Delta^0-\eps+h^{l'}\Lambda^{l'}_{Q})^{-1} + G_{\rm imp}^0 - V^{l''} L^{l''}_{Q}}.   
\end{equation}
In analogy with the previous consideration, we determine $V_{\rm MF}^l L^l_{Q}$ from the observation that $G_{\rm FLF}^{\rm MF}$ corresponds to a polarized theory, where nonlocal  fluctuations are neglected. Therefore, it should coincide with a known mean-field result. Whereas the saddle point value of the FLF~\eqref{eq:condition} within the FLF-HF scheme was determined from the polarized HF result, here  we require that~\eqref{DMFT-saddle-new} reproduces the polarized DMFT solution~\eqref{DMFT-G-new}. 
This results in the following relation
\begin{align}
V_{\rm MF}^l L^l_{Q} 
&= G_{\rm imp}^0 \, \overline{\Sigma} \, G_{\rm imp}^0,
\label{nusDMFT-new}
\end{align}
where $\overline{\Sigma}^{-1} = \delta\Sigma_{\rm imp}^{-1} + G_{\rm imp}^0$, and $\delta\Sigma_{\rm imp}$ is the difference between the non-polarized and polarized self-energies of DMFT.
This relation defines the frequency-dependent profile of the tensor quantity $L_{Q\omega}$. 

From the very beginning, the FLF $V$ and function $L_{Q}$ are introduced as a scalar product. This gives us a freedom to choose both quantities separately up to a rescaling parameter. 
For numerical calculations it is convenient to use the normalized value of the FLF imposing that $||{\bf V}_{\rm MF}^2||=1$.  
Then, substituting the result of Eq.~\eqref{nusDMFT-new} to Eq.~\eqref{lambda-DMFT-new}, one immediately finds an effective stiffness constant $J_{Q}$.
Note that, the proposed choice for the saddle point value of the FLF~\eqref{nusDMFT-new} is not unique. In principle, one can find other physical arguments to fix $V_{\rm MF}$. One more possibility that determines the saddle point is discussed in Appendix~\ref{AppendixB}. However, we find that it does not lead to a noticeable change of the result of the introduced FLF-DMFT theory.     

\section{Numerical results}

In this section we present numerical results for the Curie constant $C=\beta^{-1}\partial_h \av{s}$ for Hubbard plaquettes with periodic boundary conditions. 
Two regimes of a moderate ($U=2 t$)  and strongly ($U=8 t$) correlated system are considered. 
Results for the FLF-HF and FLF-DMFT calculations are compared to their parental approximations and to the reference lattice quantum Monte Carlo (QMC) data. At $h=0$ the susceptibility tensor $\partial_{\bf h}  {\bf m}$ is isotropic. Its diagonal component is equal to $\frac13 \partial_{h^l h^l} \ln {\cal Z}$, 
where the factor 1/3 compensates the summation over the index $l$. 
The explicit relation for the FLF-HF theory can be found using the corresponding partition function ${\cal Z}_{\rm FLF}$~\eqref{eq:Z_DMFT}.
The second derivative of the partition function then reads
\begin{align}
\label{Z2-HF-num}
\frac{\partial^{2} {\cal Z}_{\rm FLF}}{\partial h^l \partial h^l} &= 
\int \left( \left[\frac{\partial {\cal S}_{\rm FLF}[V]}{\partial h^l}\right]^2 - \frac{\partial^{2} {\cal S}_{\rm FLF}[V]}{\partial h^{l} \partial h^l} \right) e^{-{\cal S}_{\rm FLF}[V]} \,V^2 dV.
\end{align}
Derivatives at the r.h.s. of this equation are obtained numerically. 
Note that in the absence of the external field $h$, the problem becomes isotropic, and the integral over the vector field $V$ reduces to a single-variable integral over the absolute value of $V$.
FLF-DMFT calculations start with obtaining self-energies for a polarized and non-polarized DMFT solution. 
For this aim we use the exact diagonalization solver and apply $h=0.005 t$ as a small polarizing field. This allows us to obtain $J_{Q}$ and $L_{Q}$ quantities according to above described procedure.
Further calculations are performed in the same way as for the FLF-HF theory.

Let us turn to a comparison of obtained numerical results for all mentioned theories against a benchmark QMC result. 
Fig.~\ref{figure-Curie} shows an effective AFM Curie constant $C$ as a function of an inverse temperature $\beta$ for a $4\times4$ plaquette for $U=2t$ (top panels) and $U=8t$ (bottom panels). 
Left column corresponds to the HF case, and right column shows the result obtained within DMFT scheme.
The QMC data demonstrates that lowering the temperature the Curie constant first increases. 
This corresponds to a formation and softening of a collective AFM mode. At a certain point $C(\beta)$ saturates, which is clearly visible in a strongly-interacting regime. 
For very low temperatures that are not shown in the Figure, $C(\beta)$ is expected to decay as $\beta^{-1} \chi_{gs}$ with $\chi_{gs}$ being a ground state susceptibility of the system. 
 
We observe that the HF result agrees with the reference data only for a very high temperature. 
Lowering the temperature, the HF drastically overestimates the Curie constant and shows an unphysical (artificial) phase transition. 
The HF N\'eel point is indicated in Fig.~\ref{figure-Curie} by a vertical dashed line. 
We note that a significant overestimation is seen already for temperatures far above the N\'eel point. 
Thus, we find that the applicability of the HF approximation is very limited even for a moderately correlated case $U=2t$. 
Compared to the HF method, DMFT leads to a quantitatively much better result for the Curie constant. 
In particular, the N\'eel temperature predicted by DMFT in a strongly correlated regime $U=8t$ is several times lower than the one of the HF. 
However the qualitative behavior of $C(\beta)$ at low temperatures remains the same.
 
The FLF extension dramatically improves the result of both mean-field approaches. For instance, an precise account for AFM fluctuations allows to prevent a spontaneous symmetry breaking associated with the AFM ordering. 
It is important to point out that, although the FLF calculations use the mean-field data as a starting point, resulting FLF curves for $C(\beta)$ remain smooth at N\'eel temperatures predicted by bare mean-field theories. 
We find that FLF theories are in a good agreement with benchmark QMC data, especially for the FLF-DMFT calculations.
Thus, the high-temperature region, where the FLF-DMFT approximation reproduces QMC points, is remarkably larger compared to the bare DMFT case. At lower temperatures, we observe a uniform discrepancy of about $20\%$ between the FLF-DMFT and QMC results.

\begin{figure}[t!]
\includegraphics[width=0.95\linewidth]{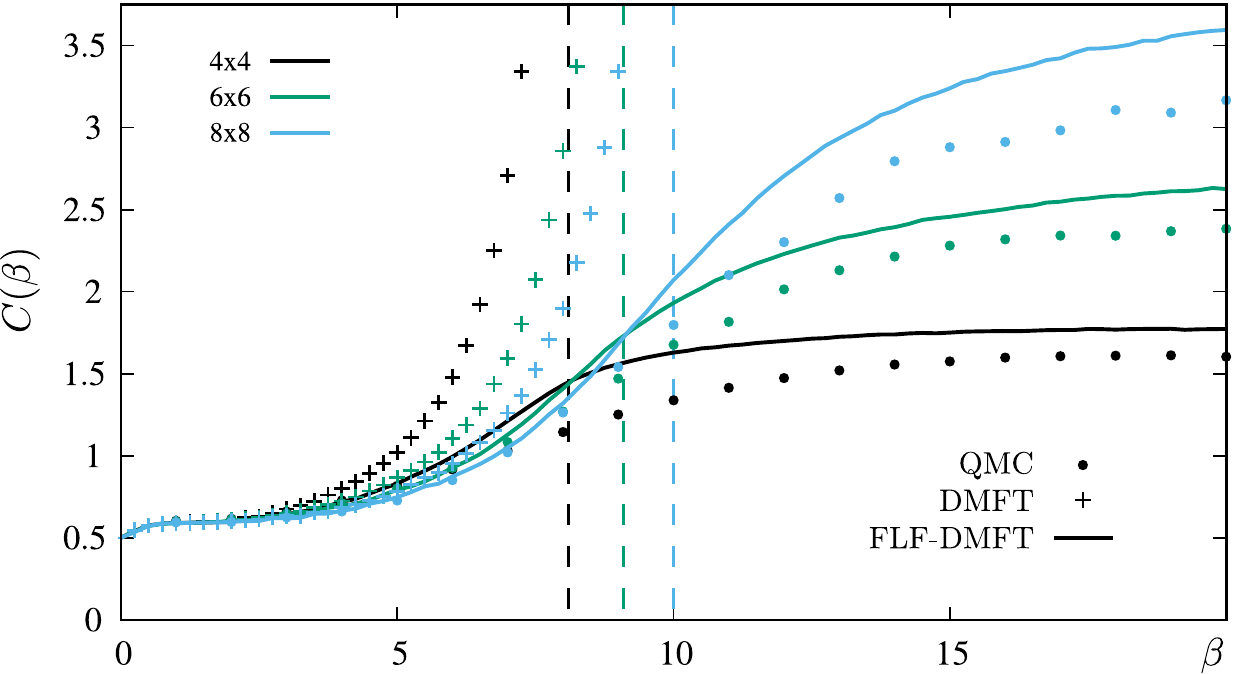} 
\caption{\label{figure-size} Curie constant $C(\beta)$ obtained for $U=2t$ for $4\times4$ (black), $6\times6$ (green), and $8\times8$ (blue) plaquettes as a function of the inverse temperature $\beta$. The result is compared for different QMC (dots), FLF-DMFT (solid line), and DMFT (crosses) approaches. Vertical dashed lines depict the DMFT N\'eel point, which depends on the size of plaquettes. } 
\end{figure}

As a next step, let us demonstrate how the FLF-DMFT theory performs for larger plaquettes containing $6\times 6$ and $8 \times 8$ lattice sites, and for a wider range of temperatures including the region way below the DMFT N\'eel point. 
Corresponding results are shown in Fig.~\ref{figure-size}. 
Here, we limit ourselves to a moderately interacting case of $U=2t$, mostly because in this regime the reference QMC data can to obtained without heavy numerical efforts. 
At the same time, we stress that correlation effects at $U=2t$ are by no means weak. 
This can be concluded from the fact that the N\'eel temperature predicted by DMFT is more than two times lower than the one of the HF theory (see Fig.~\ref{figure-Curie}). 
More elaborate studies also confirm that electron correlations become important at $U=2t$~\cite{PhysRevLett.124.117602, 2020arXiv200610769S}.
Indeed, although in this case the on-site Coulomb potential $U$ is much smaller than the bandwidth $W=8t$, the value of $U$ should rather be compared to the width of a much narrower peak formed at the Fermi level in the electronic density of states.

As can be seen from the reference QMC data presented in Fig.~\ref{figure-size}, the change in the plaquette size results in two effects. First, the initial increase of the the Curie constant upon decreasing the temperature is slower for larger lattices. This trend is especially visible for a temperature range $4\lesssim\beta\lesssim8$ and
is related to the local density of electronic states. 
Bare DMFT calculations qualitatively capture this effect. The same mechanism is responsible for a decrease of the DMFT N\'eel temperature upon increasing the plaquette size. 
The second effect is a significant increase of the Curie constant with the plaquette size below the DMFT N\'eel temperatures. However, this temperature range lies beyond the limit of applicability of DMFT. 
We emphasise, that the DMFT N\'eel point or a slightly lower temperature is also a practical limitation for diagrammatic schemes constructed around DMFT~\cite{2020arXiv200610769S}.

\begin{figure}[t!]
\includegraphics[width=0.95\linewidth]{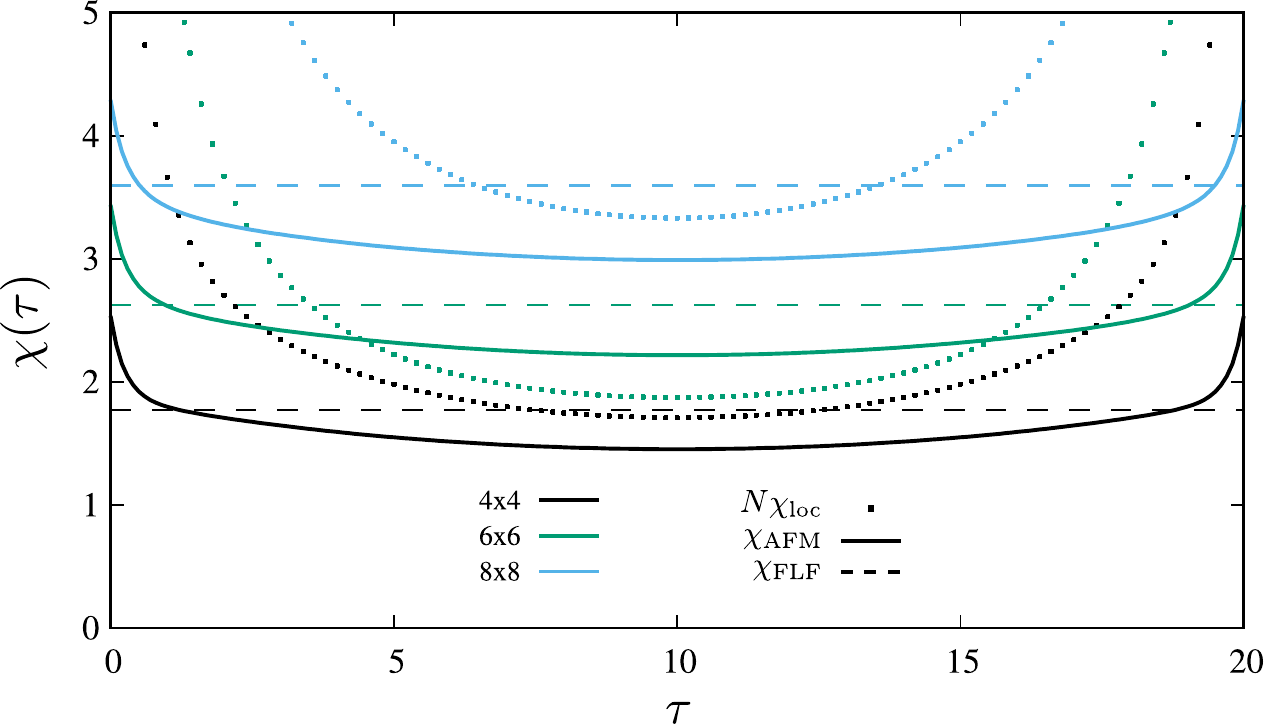} 
\caption{\label{figure-PiPivsLoc} Magnetic susceptibility $\chi(\tau)$ calculated for $4\times4$ (black), $6\times6$ (green), and $8\times8$ (blue) plaquettes as a function of the imaginary time $\tau$. Local (dots) and AFM (solid lines) susceptibilities are exact QMC results. Horizontal dashed lines correspond to FLF-DMFT approximation.} 
\end{figure}

Before switching to FLF-DMFT results, let us discuss what kind of change in the FLF data can be expected upon increasing the plaquette size.
First, for larger plaquettes the AFM mode, which is associated with the wave vector $Q=(\pi, \pi)$, affects more lattice sites and thus becomes ``more classical''. 
The fact that the FLF considers only classical fluctuations of the order parameter makes this method more appropriate for plaquettes that are not too small.
However, for very large plaquettes other spatial fluctuations of the order parameter associated with ${\bf q}\neq Q$ become important. 
This argument is confirmed by the QMC result for the imaginary time dependence of the magnetic susceptibility $\chi_{\bf q}(\tau)$. 
In Fig.~\ref{figure-PiPivsLoc} the AFM component of the susceptibility obtained for ${\bf q} = Q$ (solid line) is compared to the local susceptibility summed over all wave vectors $N \chi_{loc}(\tau) = \sum_{\bf q} \chi_{\bf q}(\tau)$ (dotted line). 
The result is presented for $\beta=20$, which is well below the DMFT N\'eel point. 
The FLF-DMFT predictions are indicated by dashed horizontal lines. 
First of all we observe that, except for a high-energy tails seen near $\tau=0$ and $\tau=\beta$, the AFM susceptibility indeed shows a weak $\tau$-dependence for all considered lattices. 
Also, we find that at $\tau=\beta/2$ the value of $N \chi_{loc}$ is very close to $\chi_{Q}$. 
This proves that the low-energy collective fluctuations are dominated by a single AMF mode associated with the $Q=(\pi,\pi)$ and $\Omega=0$ channel, which justifies the main idea of the FLF approach. 
It can be seen that deviation of $N \chi_{loc}(\tau)$ from $\chi_{Q}(\tau)$ grows with the lattice size indicating that contributions of other ${\bf q}\neq Q$ fluctuations become more important. 
This analysis suggests that FLF-DMFT approach is best suited for medium plaquettes.

Fig.~\ref{figure-size} confirms the above reasoning. 
Indeed, among all plaquettes the best numerical accuracy of the FLF-DMFT is observed for the $6 \times 6$ system.
For the $4 \times 4$ plaquette, the biggest deviation from the benchmark result is observed near the DMFT N\'eel point, where a Mexican-hat potential starts to form. 
It can be concluded that quantum fluctuations of the order parameter are particularly important in this regime. 
On the other hand, the biggest overestimation of the Curie constant for the $8\times 8$ plaquette is observed at low temperatures.
As we have discussed above, such an overestimation is associated with the neglection of uncorrelated spatial fluctuations of the order parameter. It can be seen, that the FLF-DMFT method in its present form reduces to bare DMFT in the limit of an infinite lattice, because the saddle point estimation of the integral over the FLF becomes exact. Since the DMFT predicts a divergence in the Curie constant, the FLF-DMFT is expected to show a larger overestimation of $C$ for bigger plaquettes.

Nevertheless, the overall performance of the FLF-DMFT approach and its agreement with the reference data is found to be quite satisfactory.
In fact, below the DMFT N\'eel point $T^{\rm DMFT}_{N}$ the discrepancy between the FLF-DMFT and reference QMC data does not exceed 20\% for all considered plaquettes. Moreover, this difference remains almost unchanged even for very low temperatures $T\simeq0.5\,T^{\rm DMFT}_{N}$. 
Note that this regime of temperatures is known to be extremely difficult for numerical calculations, because even the most advanced theoretical approaches are not able to provide reasonable results below ${T\simeq0.75\,T^{\rm DMFT}_{N}}$~\cite{2020arXiv200610769S}.
In this regard, we can qualitatively compare the result of the FLF-DMFT theory for the $8\times8$ plaquette with the results of existing theories obtained for the infinite lattice. One finds that enlarging the lattice size from $8\times 8$ to infinity, the DMFT N\'eel temperature decreases from $T=0.10t$ to $T=0.08t$. Then, for the same value of the on-site Coulomb potential $U=2t$ considered in our work, we observe that the exact diagrammatic Monte-Carlo (DiagMC) methods~\cite{PhysRevD.24.2278, PhysRevLett.81.2514} do not converge below ${T=0.83\,T^{\rm DMFT}_{N}}$. The dual fermion~\cite{PhysRevB.77.033101, PhysRevB.79.045133, PhysRevLett.102.206401, BRENER2020168310} and the dual boson~\cite{Rubtsov20121320, PhysRevB.90.235135, PhysRevB.93.045107, PhysRevB.94.205110, PhysRevB.100.165128} approaches quantitatively agree with the DiagMC result for magnetic susceptibility, but are limited to the temperature range $T\gtrsim0.78\,T^{\rm DMFT}_{N}$ (see Fig.\,13 of the Ref.~\onlinecite{2020arXiv200610769S}). Diagrammatic extensions of DMFT such as the dynamical vertex approximation (D$\Gamma$A)~\cite{PhysRevB.75.045118, PhysRevB.80.075104} and the TRILEX theory~\cite{PhysRevB.92.115109, PhysRevB.93.235124, PhysRevLett.119.166401} can perform calculations for a bit lower temperature $T\simeq0.75\,T^{\rm DMFT}_{N}$ and $T\simeq0.63\,T^{\rm DMFT}_{N}$, respectively.
However, at the DMFT N\'eel point both theories deviate from the exact result already by $10-15\%$, and this accuracy rapidly decays to $30-50\%$ when lowering the temperature.
In contrast, the FLF-DMFT result for the $8\times8$ plaquette presented in Fig.~\ref{figure-size} shows a uniform discrepancy of about $10-15\%$ down to the lowest considered temperature, that is ${T=0.5\,T_{N}^{\rm DMFT}}$.
This illustrates a conceptual advantage of the FLF method that exactly accounts for strong AFM fluctuations, which cause problems for all existing diagrammatic extensions of DMFT.

\section{Conclusions and outlook}

We have demonstrated that an accurate account for fluctuations of soft collective modes is crucially important for a correct description of a low-temperature behavior of correlated electronic systems. 
For this aim we have introduced and tested a novel fluctuating local field technique, which is capable to handle these collective modes in a wide temperature range, including a strongly nonlinear regime of fluctuations.
Compared to an exact QMC solution for half-filled Hubbard plaquettes, the FLF-DMFT scheme shows a quantitatively good result for Curie constant. 
The reason is that FLF-DMFT explicitly considers collective AFM fluctuations in addition to local correlations accounted by a bare DMFT approach.

In the present paper ve have benchmarked the FLF theory for a half-filled Hubbard plaquettes with emerging classical fluctuations of a single collective AFM mode. 
This system has been chosen because of its conceptual simplicity, availability of the numerically exact reference data, and, simultaneously, for a clear lack of existing approximations that can provide a satisfactory solution at low temperature. 
The FLF-DMFT approach shows an accuracy of about $20\%$ or better for all considered regimes including the temperature twice lower than the DMFT N\'eel point. 
We consider these results rather promising, bearing in mind that the introduced computational scheme does not require significant numerical efforts beyond DMFT calculations. 

In its present form the FLF method can be straightforwardly applied to a wide range of finite correlated systems that exhibit strong collective (charge, spin, etc.) fluctuations with dominant single or few collective modes. 
In particular, molecular magnets can be seen as attractive candidates for the first realistic application of the FLF theory.
Technically, leading collective modes can be determined from the instabilities and/or multiple solutions that arise from the mean-field consideration. 
For example, one can propose a single-stripe solution for the $L_x \times L_y$ Hubbard plaquette away from the half-filling by imposing fluctuations in the $(\pi-2 \pi/L_{x}, \pi)$ channel. 
Of course, a quantitative accuracy of this scheme requires an additional consideration.
On the other hand, modelling of large 2D lattices would require an extension of the theory towards an inclusion of other than AFM spatial fluctuating modes. 
This problem can be addressed, for example, by modelling the infinite lattice within a cluster scheme, and using the FLF method as a cluster solver. However, this would require an additional study whether the FLF can produce sufficiently accurate data for the Green's function. 
A possible improvement of the numerical accuracy of the FLF-DMFT method can be achieved by using a better reference system. 
Here, we can benefit from the fact that the introduced scheme is based on the dual fermion ideology and thus, is not restricted only to the single-site impurity problem of DMFT~\cite{BRENER2020168310}.
Then, the use of a small, e.g. $2\times 2$ cluster as the reference system will allow to consider short-range correlations exactly, whereas the collective AFM fluctuations can be accounted for by the fluctuating field. 
Importantly, the FLF method can be formulated for all possible cluster schemes, such as the free-standing cluster or the cluster with periodic boundary conditions~\cite{RevModPhys.77.1027, RevModPhys.78.865}. 
For this purpose, the impurity Green's function and hybridisation in the FLF-DMFT equations should be simply replaced by corresponding quantities of the cluster scheme.

In a broader context, it is worth to relate our method to other known approached that handle strong fluctuations in quantum systems. 
Whenever leading fluctuating degrees of freedom can be clearly isolated, constrained calculations are widely used. 
For example, the analysis of molecular conformations in quantum chemistry is essentially based on the estimation of the electronic energy for a constrained atomic configuration~\cite{doi:10.1021/cr200137a}. 
However, there is no good way to introduce a constrain for collective degrees of freedom -- as, for instance, to keep the total magnetization fixed while allowing for single-electron excitations. 
Instead, in our work we have introduced an additional variable, namely the fluctuating field $V$, that affects the {\it average} value of the order parameter. 
It resembles a seminal approach by Wheeler et al. known as the generator coordinate method~\cite{Wheeler1, Wheeler2}. It is widely used in the field of nuclear physics, and also has proved its efficiency in a number of developments including quantum chemistry~\cite{CapelleDFT}. 
This variational approach works with a set of configurations that differ by the value of the field acting on the system.  
In practice, only few configurations are considered, otherwise the method becomes computationally expensive.  
This is the most obvious difference from the FLF theory, where the integral is taken over a continuously varied fluctuating field, which in particular allows us to respect the spin-rotation symmetry. 
Another prominent link can be established to the contemporary field theories based on the functional renormalization group (fRG)~\cite{polonyi2003lectures}. 
This method relies on the generating functional ${W(V)=\ln \int D[\phi] \exp\{-S[\phi] + V \phi\}}$,  where the action $S[\phi]$ describes the nonlinear bosonic field theory. 
Thus, the FLF-HF can be seen as a generalization of the  previously known methods~\cite{Wheeler1, Wheeler2, CapelleDFT, polonyi2003lectures}.
On the other hand, the unique advantage of the FLF-DMFT scheme over other known approaches is that it allows for a simultaneous unperturbative treatment of the on-site correlations together with spatial fluctuations of the ``global'' order parameter.

\begin{acknowledgments}
The authors are  very grateful to Maria Bandelmann for the help with graphics.
The work of E.A.S. is supported by the Russian Science Foundation Grant 18-12-00185.
A.I.L. acknowledges the support by the Cluster of Excellence ``Advanced Imaging of Matter'' of the Deutsche Forschungsgemeinschaft (DFG) - EXC 2056 - Project No. ID390715994.
\end{acknowledgments}

\appendix

\onecolumngrid

\section{Variational derivation of the FLF-HF method}
\label{app:Variation}

In this section we present another way of deriving the FLF-HF theory.
As we discuss in the main text, the FLF action
\begin{align}
{\cal S}_{\rm FLF} [\cd, c, V] 
&= - \cd_{1} {\cal G}^{-1}_{12} c_{2} - V^l M^l_{Q} + \frac12\frac{\beta N}{J_{Q}} {\bf V}^2.
\label{eq:FLF}
\end{align}
is Gaussian in terms of the introduced fluctuating field $V$. Therefore, this field can be integrated out exactly leading to an effective interacting fermionic problem 
\begin{align}
{\cal S}^{*}_{\rm FLF} [\cd, c] 
&= - \cd_{1} {\cal G}^{-1}_{12} c_{2} - \frac12 \frac{J_{Q}}{\beta N} {\bf M}^2,
\label{eq:mapping}
\end{align}
where ${\bf M}^2 = M^l_{Q} M^l_{Q}$.
The initial problem
\begin{align}
\label{Hubbard}
{\cal S}[\cd,c] &= - \cd_{1} {\cal G}^{-1}_{12} c^{\phantom{*}}_{2}  + U 
\left(n_{j\tau\uparrow}-\frac{1}{2}\right)\left(n_{j\tau\downarrow}-\frac{1}{2}\right).  
\end{align}
can be mapped onto this new action~\eqref{eq:mapping} using Peierls-Feynman-Bogoliubov variational principle~\cite{PhysRev.54.918, bogolyubov1958variation, feynman1972statistical} for the functional
\begin{align}
\tilde{\cal F}(J_{Q}) = {\cal F}^{*}_{\rm FLF} + (\beta{}N)^{-1}\av{{\cal S}[\cd,c] - {\cal S}^{*}_{\rm FLF}[\cd,c]}_{*}.
\end{align}
Here, ${\cal F}^{*}_{\rm FLF} = -(\beta{}N)^{-1}\ln{\cal Z}^{*}_{\rm FLF}$ is the free energy of the new problem~\eqref{eq:mapping}, and $\av{\ldots}_{*}$ denotes averaging with respect to the partition function ${\cal Z}^{*}_{\rm FLF}$. The optimal value for the parameter $J_{Q}$ can be found from the following condition $\partial_{J_{Q}}\tilde{\cal F}(J_{Q})=0$ that minimizes the energy. One finds that
\begin{align}
\partial_{J_{Q}}{\cal F}^{*}_{\rm FLF} &= -\frac{1}{\beta{}N}\frac{1}{{\cal Z}^{*}_{\rm FLF}} \int D[c^{\dagger},c] \, \frac{1}{2\beta{}N} {\bf M}^{2}\, e^{-{\cal S}^{*}_{\rm FLF}[c^{\dagger},c]} 
= -\frac{1}{2(\beta{}N)^2} \av{ {\bf M}^{2}}_{*}.
\end{align}
and
\begin{align}
\frac{1}{\beta{}N}\partial_{J_{Q}}\av{{\cal S}[\cd,c] - {\cal S}^{*}_{\rm FLF}[\cd,c]}_{*} 
&= \frac{U}{\beta{}N}\partial_{J_{Q}} \av{\left(n_{j\tau\uparrow}-\frac{1}{2}\right)\left(n_{j\tau\downarrow}-\frac{1}{2}\right)}_{*} + \partial_{J_{Q}}\av{\frac{J_{Q}}{2(\beta{}N)^2} {\bf M}^{2}}_{*} \notag\\
&= \frac{U}{\beta{}N}\partial_{J_{Q}} \av{\left(n_{j\tau\uparrow}-\frac{1}{2}\right)\left(n_{j\tau\downarrow}-\frac{1}{2}\right)}_{*} 
+ \frac{1}{2(\beta{}N)^2}  \av{{\bf M}^{2}}_{*} + \frac{J_{Q}}{2(\beta{}N)^2}  \partial_{J_{Q}}\av{{\bf M}^{2}}_{*}.
\end{align}
Therefore, the final expression for the parameter $J_{Q}$ reads
\begin{align}
J_{Q} = - 2U\beta{}N \frac{ \partial_{J_{Q}} \av{ \left(n_{j\tau\uparrow}-\frac{1}{2}\right)\left(n_{j\tau\downarrow}-\frac{1}{2}\right)}_{*} }{  \partial_{J_{Q}}\av{{\bf M}^{2}}_{*} }.
\label{eq:lambda1}
\end{align}
To evaluate this expression, one can transform the average over the action~\eqref{eq:mapping} to the average over the FLF action~\eqref{eq:FLF} as
\begin{align}
\av{O[c^{\dagger},c]}_{*} &= \frac{1}{{\cal Z}^{*}_{\rm FLF}} \int D[c^{\dagger},c] \, O[c^{\dagger},c]\, e^{-{\cal S}^{*}_{\rm FLF}[c^{\dagger},c]} \notag\\
&= \frac{1}{{\cal Z}_{\rm FLF}} \int D[c^{\dagger},c]\, d^{3}V \, O[c^{\dagger},c]\, e^{-{\cal S}_{\rm FLF}[c^{\dagger},c,V]} \notag\\
&= \frac{1}{{\cal Z}_{\rm FLF}} \int d^{3}V \, e^{-\frac12\frac{\beta N}{J_{Q}}{\bf V}^2} \int D[c^{\dagger},c]\,O[c^{\dagger},c]\, e^{\cd_{1} {\cal G}^{-1}_{12} c_{2} + V^l s^l_{j\tau}} \notag\\
&= \frac{1}{{\cal Z}_{\rm FLF}} \int d^{3}V \, e^{-\frac12\frac{\beta N}{J_{Q}}{\bf V}^2} \frac{\int D[c^{\dagger},c]\,O[c^{\dagger},c]\, e^{\cd_{1} {\cal G}^{-1}_{12} c_{2} + V^l s^l_{j\tau}}}{\int D[c^{\dagger},c]\,e^{\cd_{1} {\cal G}^{-1}_{12} c_{2} + V^l s^l_{j\tau}}}\int D[c^{\dagger},c]\,e^{\cd_{1} {\cal G}^{-1}_{12} c_{2} + V^l s^l_{j\tau}} \notag\\
&= \frac{1}{{\cal Z}_{\rm FLF}} \int D[c^{\dagger},c] \, d^{3}V \, e^{-{\cal S}_{\rm FLF}[c^{\dagger},c,V]} \av{O[c^{\dagger},c]}_{{\rm FLF}_{e}}
\end{align}
where $\av{\ldots}_{{\rm FLF}_{e}}$ denotes the average over the fermionic part of the FLF action~\eqref{eq:FLF}.
Note that $\av{O[c^{\dagger},c]}_{{\rm FLF}_{e}}$ depends on the field $V$, but does not depend on fermionic variables $c^{(\dagger)}$. 
Thus, one can integrate out fermionic degrees of freedom and get
\begin{align}
\av{O[c^{\dagger},c]}_{*} 
&= \frac{1}{\int d^3V\,e^{-{\cal S}_{\rm FLF}[V]}} \int d^3V\,\av{O[c^{\dagger},c]}_{{\rm FLF}_{e}}\,e^{-{\cal S}_{\rm FLF}[V]}
= \av{\av{O[c^{\dagger},c]}_{{\rm FLF}_c}}_{{\rm FLF}_{V}}
\end{align}
where $\av{\ldots}_{{\rm FLF}_{V}}$ denotes the average over the action ${\cal S}_{\rm FLF}[V]$, which is given by 
\begin{equation}
\label{Fnu}
{\cal S}_{\rm FLF}[V]=-\ln \det \left[{\cal G}^{-1} + V^l \Lambda^l_{Q}\right] + \frac12\frac{\beta N}{J_{Q}} {\bf V}^2.
\end{equation}
and depends only on the field $V$.

Now, one can calculate averages that enter the Eq.~\eqref{eq:lambda1} as
\begin{align}
\partial_{J_{Q}} \av{ \left(n_{j\tau\uparrow}-\frac{1}{2}\right)\left(n_{j\tau\downarrow}-\frac{1}{2}\right)}_{*} &=
\partial_{J_{Q}} \av{\av{ \left(n_{j\tau\uparrow}-\frac{1}{2}\right)\left(n_{j\tau\downarrow}-\frac{1}{2}\right)}_{{\rm FLF}_c}}_{{\rm FLF}_{V}}.
\end{align}
The fermion part of the FLF action is Gaussian, so the first average is equal to
\begin{align}
\av{ \left(n_{j\tau\uparrow}-\frac{1}{2}\right)\left(n_{j\tau\downarrow}-\frac{1}{2}\right)}_{{\rm FLF}_c} 
= 
-\frac{\beta{}N}{4} {\bf m}^2_{{\rm FLF}_{c}},
\end{align}
and the result is
\begin{align}
\partial_{J_{Q}} \av{ \left(n_{j\tau\uparrow}-\frac{1}{2}\right)\left(n_{j\tau\downarrow}-\frac{1}{2}\right)}_{*} &=
-\frac{\beta{}N}{4} \partial_{J_{Q}} \av{{\bf m}^2_{{\rm FLF}_{c}}}_{{\rm FLF}_{V}},
\end{align}
The average over fermionic degrees of freedom of the denominator of Eq.~\eqref{eq:lambda1} reads
\begin{align}
\label{eq:AvSS}
\av{s^{2}_{\rm AFM}}_{{\rm FLF}_{c}} &= 
(\beta{}N)^2 {\bf m}^2_{{\rm FLF}_{c}}  -  \Lambda^{l}_{i}\,{\cal G}_{ij}(\tau,\tau')\,\Lambda^{l}_{j}\,{\cal G}_{ji}(\tau',\tau)
\end{align}
Last term in this equation is much smaller than the first one. So, the result for the parameter $J_{Q}$ reduces to 
\begin{align}
J_{Q} = \frac{U}{2}\frac{\partial_{J_{Q}} \av{{\bf m}^2_{{\rm FLF}_{c}}}_{{\rm FLF}_{V}} }{(\beta{}N)^{-2}\,\partial_{J_{Q}} \av{\av{{\bf M}^{2}}_{{\rm FLF}_{c}}}_{{\rm FLF}_{V}}} \simeq \frac{U}{2}.
\label{eq:lambda2}
\end{align}

\section{Path-integral derivation of the FLF-DMFT theory}
\label{AppendixB}

In this section we present a detailed derivation of the FLF-DMFT theory and discuss different possibilities to introduce a fluctuating field and a saddle-point condition.  
First, we start with the lattice action~\eqref{Hubbard} and explicitly isolate the local impurity problem of DMFT ${\cal S}^{(j)}_{\rm imp}[\cd,c]$
\begin{align}
{\cal S}[\cd,c] &= 
\sum_{j}{\cal S}^{(j)}_{\rm imp}[\cd,c] + \cd_{\kv\omega\sigma} \epsilon^{\sigma\sigma'}_{\kv\qv\omega} c^{\phantom{\dagger}}_{\kv+\qv,\omega\sigma'},
\end{align}
where we introduce $\epsilon^{\sigma\sigma'}_{\kv\qv\omega} = \varepsilon^{\phantom{\dagger}}_{\kv}\delta_{\sigma\sigma'}\delta_{\qv,0} - \Delta^{\sigma\sigma'}_{\qv\omega} - \sigma^{l}_{\sigma\sigma'}h^{l}\delta_{\qv,Q}$. In order to account for local correlation effects exactly, we integrate out the impurity problem following the idea of the dual fermion theory~\cite{PhysRevB.77.033101}. For this aim, we first perform Hubbard-Stratonovich transformation over the nonlocal part of the action
\begin{align}
&\exp\left\{ \cd_{\kv\omega\sigma} \left[-\epsilon^{\sigma\sigma'}_{\kv\qv\omega}\right] c^{\phantom{\dagger}}_{\kv+\qv,\omega\sigma'}\right\} = \det\left[-\epsilon^{\sigma\sigma'}_{\kv\qv\omega}\right] 
\int D[f^{\dagger},f] \exp\left\{-\left( f^{\dagger}_{\kv\omega\sigma}\left[-\epsilon^{\sigma\sigma'}_{\kv\qv\omega}\right]^{-1}f^{\phantom{\dagger}}_{\kv+\qv,\omega\sigma'} + c^{\dagger}_{\kv\omega\sigma}f^{\phantom{\dagger}}_{\kv\omega\sigma} + f^{\dagger}_{\kv\omega\sigma}c^{\phantom{\dagger}}_{\kv\omega\sigma}\right)\right\}.
\end{align}
Now, initial fermionic variables $c^{(\dagger)}$ can be integrated out with respect to the impurity action. This results in the following form of the partition function 
\begin{align}
{\cal Z} &= \det\left[-\epsilon^{\sigma\sigma'}_{\kv\qv\omega}\right] \int D[f^{\dagger},f] \exp\left\{-f^{\dagger}_{\kv\omega\sigma}\left[-\epsilon^{\sigma\sigma'}_{\kv\qv\omega}\right]^{-1}f^{\phantom{\dagger}}_{\kv+\qv,\omega\sigma'} \right\}
\int D[c^{*},c] \exp\left\{-\sum_{j}{\cal S}^{(j)}_{\rm imp}[\cd,c] -  \cd_{\kv\omega\sigma}f^{\phantom{\dagger}}_{\kv\omega\sigma} - f^{\dagger}_{\kv\omega\sigma} c^{\phantom{\dagger}}_{\kv\omega\sigma} \right\} \notag\\
&= \det\left[-\epsilon^{\sigma\sigma'}_{\kv\qv\omega}\right] {\cal Z}_{\rm imp} \int D[f^{\dagger},f] \exp\left\{-\tilde{\cal S}[f^{\dagger},f]\right\}
\label{eq:Z_DF}
\end{align}
with the dual fermion action 
\begin{align}
\tilde{\cal S}[f^{\dagger},f] = - f^{\dagger}_{1} \tilde{\cal G}^{-1}_{12} f^{\phantom{\dagger}}_{2} + W[f^{\dagger},f].
\label{eq:S_DF}
\end{align}
Here, $\tilde{G}$ is the bare dual Green's function that can be found from the following relation
$
\tilde{\cal G}^{-1} = \epsilon^{-1} - G_{\rm imp},
$
where $G_{\rm imp}$ is the exact Green's function of the impurity problem. The interaction part $W[f^{\dagger},f]$ contains all possible local fermion-fermion vertex functions~\cite{PhysRevB.77.033101}. Neglecting the interaction, the theory reproduces the DMFT result. To illustrate this, let us perform a back transformation to the initial fermion variables explicitly
\begin{align}
\exp\left\{-f^{\dagger}_{\kv\omega\sigma} \left[-\epsilon^{\sigma\sigma'}_{\kv\qv\omega}\right]^{-1}f^{\phantom{\dagger}}_{\kv+\qv,\omega\sigma'} \right\} = 
-\det\left[-\epsilon^{\sigma\sigma'}_{\kv\qv\omega}\right]^{-1}
\int D[c^{\dagger},c] \exp\left\{ \cd_{\kv\omega\sigma} \left[-\epsilon^{\sigma\sigma'}_{\kv\qv\omega}\right] c^{\phantom{\dagger}}_{\kv+\qv,\omega\sigma'} + \cd_{\kv\omega\sigma}f^{\phantom{\dagger}}_{\kv\omega\sigma} + f^{\dagger}_{\kv\omega\sigma}c^{\phantom{\dagger}}_{\kv\omega\sigma}\right\}.
\end{align}
The total partition function then reads
\begin{align}
{\cal Z}
&= - {\cal Z}_{\rm imp} \int D[\cd,c]
\exp\left\{ \cd_{\kv\omega\sigma} \left[-\epsilon^{\sigma\sigma'}_{\kv\qv\omega}\right] c^{\phantom{\dagger}}_{\kv+\qv,\omega\sigma'} \right\}
\int D[f^{\dagger},f] \exp\left\{  \cd_{\kv\omega\sigma}f^{\phantom{\dagger}}_{\kv\omega\sigma} + f^{\dagger}_{\kv\omega\sigma}c^{\phantom{\dagger}}_{\kv\omega\sigma} - 
f^{\dagger}_{\kv\omega\sigma} G^{\rm imp}_{\omega\sigma\sigma'} f^{\phantom{\dagger}}_{\kv\omega\sigma'} \right\} \notag\\
&= -{\rm det}\left[G_{\rm imp}\right] {\cal Z}_{\rm imp} \int D[\cd,c]
\exp\left\{ -{\cal S}_{\rm DMFT}[\cd,c] \right\},
\end{align}
where the DMFT action is
\begin{align}
{\cal S}_{\rm DMFT}[\cd, c] = -\cd_1 \left[G_{\rm DMFT}\right]^{-1}_{12}  c^{\phantom{\dagger}}_2,
\label{eq:S_DMFT}
\end{align}
and the DMFT Green's function is $G_{\rm DMFT} = \left[G_{\rm imp}^{-1} - \epsilon \right]^{-1}$.

Fluctuating local field can be, in principle, introduced for the non-polarized DMFT problem~\eqref{eq:S_DMFT}. Then, as discussed in the main text, fluctuations will be associated with the hybridization function $\Delta$ that explicitly enters $G^{-1}_{\rm DMFT}$ through $\epsilon$, and the FLF-DMFT action would look like
\begin{align}
{\cal S}_{\rm FLF}[\cd, c,V] = -\cd_1 \left[G^{-1}_{\rm DMFT} + V^{l} L^{l}_{Q} \right]_{12} c^{\phantom{\dagger}}_2  + \frac12\frac{\beta{}N}{J_{Q}} {\bf V}^2.
\label{eq:S_FLF-DMFT}
\end{align}
Following the idea of the FLF-HF theory, the saddle-point value of the fluctuating field $V_{\rm MF}$ can be fixed by a polarized solution of the DMFT. Using that the DMFT Green's function fulfills the Dyson equation $G^{-1}_{\rm DMFT} = {\cal G}^{-1} - \Sigma_{\rm imp}$, this results in the following condition
\begin{align}
V^{l}_{\rm MF}L^{l}_{Q} =   \delta\Sigma_{\rm imp},
\end{align}
where $\delta\Sigma_{\rm imp}$ is the difference between the non-polarized and polarized self-energies of DMFT.

However, we find more convenient to introduce the FLF for the Gaussian part of the non-polarized dual fermion problem~\eqref{eq:S_DF}. Then, the FLF action in the dual space is 
\begin{align}
\tilde{\cal S}_{\rm FLF}[f^{\dagger},f,V] = - f^{\dagger}_1 \left[\tilde{\cal G}^{-1}_{12} + V^{l}L^{l}_{12}\right] f^{\phantom{\dagger}}_2 + \frac12\frac{\beta{}N}{J_{Q}} {\bf V}^2.
\label{eq:S_DF}
\end{align}
As one can observe, this expression is very similar to the FLF-HF action~\eqref{eq:FLF}: the fluctuating field $V$ is linearly coupled to a collective spin degree of freedom. In this way, we construct a direct analogue of the HF-FLF theory, but using dual variables instead of original ones. An advantage of working in the dual space is that the introduced change of variables allows to consider local correlations exactly.
Therefore, the theory based on Eq.~\eqref{eq:S_DF} accounts for both the local correlations, wich are neglected in HF approach, and collective fluctuations. 

Proceeding with the mentioned analogy, the saddle-point value of the field $V_{\rm MF}$ can be found from the Hartree-Fock diagram for the dual self-energy $\tilde{\Sigma}^{\sigma\sigma'}_{\qv\omega}$. The latter can be obtained from a convolution of the local connected two-particle Green's function $G^{(2)}_{\rm imp}$, which is contained in the interaction part $W[f^{\dagger},f]$ of the dual action, with the dual Green's function. Note that, in the presence of a nonzero field ${\bf h}$, this self-energy is nonzero, because it cannot be excluded by the non-polarized DMFT self-consistency condition~\cite{PhysRevB.77.033101}. Thus, we get
\begin{align}
V^{l}_{\rm MF} L^{l}_{Q} = -\tilde{\Sigma}.
\label{eq:SPcond}
\end{align}
The corresponding lattice action can be found using the back transformation to original fermion variables introduced above. This results in the Eq.~(20) of the main text.

It can be shown, that the dual Hartree-Fock self-energy $\tilde\Sigma$ can be connected to the polarized part of the impurity self-energy $\delta\Sigma_{\rm imp}$.  
Let us consider a fully converged polarized DMFT solution with the hybridization function $\Delta$ and Green's function $G_{\rm imp}$. One can assume that the non-polarized solution can be seen as a small deviation from the polarized one leading to
$\Delta^0=\Delta + \delta \Delta, G^0_{\rm imp}=G_{\rm imp}+\delta G_{\rm imp}$. We also introduce a corresponding change to the impurity self-energy $\delta \Sigma_{\rm imp} = -\delta(G_{\rm imp}^{-1}+\Delta)$. 
Calculating the variation of the r.h.s. of this relation explicitly, one finds that 
\begin{align}
\label{dSigma}
    \delta \Sigma_{\rm imp}=G_{\rm imp}^{-2} G^{(2)}_{\rm imp} \delta \Delta,
\end{align}
Now, let us calculate the variation of the local part of the dual Green's function
\begin{align}
\delta \sum_\kv \tilde G_{\kv} &= 
\delta \left(\sum_\kv\left[\epsilon^{-1} - G_{\rm imp} - \tilde \Sigma \right]^{-1}\right)\notag\\
&=\sum_{\kv} \tilde{G}^2_{\kv}\left[ G^2_{\rm imp}\left(\delta\Delta+\delta\Sigma_{\rm imp} \right) - \epsilon^{-2}\delta\Delta + \delta\tilde{\Sigma} \right] \notag\\
&=\sum_{\kv} \tilde{G}^2_{\kv}\left[ G^2_{\rm imp}\delta\Sigma_{\rm imp} + \delta\tilde{\Sigma} - \delta\Delta \tilde{G}^{-2}_{\kv} - 2\delta\Delta G_{\rm imp} \tilde{G}^{-1}_{\kv} \right] \notag\\
&= -\delta\Delta + \sum_{\kv} \tilde{G}^2_{\kv}\left[ G^2_{\rm imp}\delta\Sigma_{\rm imp} + \delta\tilde{\Sigma} \right],
\end{align}
where we exploited the DMFT self-consistency condition $\sum_\kv \tilde{G}_{\kv}=0$ for the last transformation.
The Hartree-Fock approximation for the dual self-energy reads
\begin{align}
\delta \tilde{\Sigma} = G^{(2)}_{\rm imp} \delta \sum_\kv \tilde{G}_{\kv}.
\end{align}
Taking Eq.~\eqref{dSigma} in to account, we get
\begin{align}
\delta \tilde{\Sigma} + G^2_{\rm imp}\delta\Sigma_{\rm imp} = \sum_{\kv} \tilde{G}^2_{\kv}G^{(2)}_{\rm imp}\left[ G^2_{\rm imp}\delta\Sigma_{\rm imp} + \delta\tilde{\Sigma} \right].
\end{align}
The solution of this equation is 
\begin{align}
\delta \tilde{\Sigma} = -G_{\rm imp}^2 \delta{\Sigma}_{\rm imp},
\label{eq:Saddle-point-dual}
\end{align}
The fact that the polarized DMFT solution corresponds to a non-polarized dual fermion result with the Hartree-Fock diagram is visible after we rewrite the polarized DMFT Green's function as
\begin{align}
G^{-1}_{\rm DMFT} &= G^{-1}_{\rm imp} - \epsilon \notag\\
&= G^{0~-1}_{\rm imp} + \delta\Sigma_{\rm imp} - \epsilon^0 \notag\\
&\simeq \left[G^{0}_{\rm imp} - G^{0}_{\rm imp}\delta\Sigma_{\rm imp}G^{0}_{\rm imp}\right]^{-1} - \epsilon^0,
\end{align}
where quantities with ``0'' index represent a non-polarized solution. This equation reproduces the exact relation between lattice and dual Green's functions if the $\delta\tilde{\Sigma} = -G^{0}_{\rm imp}\delta\Sigma_{\rm imp}G^{0}_{\rm imp}$ is the self-energy of the dual problem~\cite{PhysRevB.77.033101}.  
Thus, we have shown that the FLF-DMFT method described in the main text can be seen as an FLF-HF-like theory developed for dual variables. The saddle-point result (Eq.~(24) in the main text) in this formulation coincides (at least in the leading order) with the HF solution for dual variables providing again a full analogy with Eqs.~(11) and~(15) of the main text.

Finally, as we discuss in the main text, there exists another possibility to fix the saddle-point value for the fluctuating field $V_{\rm MF}$. Instead of equating the saddle-point result for the lattice Green's function (Eq.~(22) in the main text) to the polarized DMFT Green's function (Eq.~(19) in the main text), one can compare partition function of both problems using the relation~(18) of the main text. Then, the partition function of the polarized DMFT solution is
\begin{align}
{\cal Z}^{\phantom{1}}_{\rm DMFT} &= {\cal Z}^{\phantom{1}}_{\rm imp} \det G^{\phantom{1}}_{\rm imp} \det G^{-1}_{\rm DMFT} \notag\\
&\simeq {\cal Z}^{0}_{\rm imp} \det G^{0}_{\rm imp} \det G^{-1}_{\rm DMFT}.
\end{align}
The partition function of the dual fermion method is given by Eq.~\eqref{eq:Z_DF} and for the saddle-point Green's function~\eqref{eq:S_DF} reads
\begin{align}
{\cal Z}_{\rm DF} &= {\cal Z}^0_{\rm imp} \det\epsilon_0 \det\left[\tilde{\cal G}^{-1}_0 + V^{l}_{\rm MF} L^{l}_{Q} \right].    
\end{align}
Equating these two partition functions ${\cal Z}_{\rm DMFT} = {\cal Z}_{\rm DF}$, one gets 
\begin{align}
V^{l}_{\rm MF} L^{l}_{Q} &= \sum_{\kv}\epsilon^{-1}_0 G^{0}_{\rm imp} \delta\Sigma_{\rm imp} 
= \sum_{\kv} \left(\tilde{\cal G}^{-1}_0 + G^{0}_{\rm imp} \right)G^{0}_{\rm imp} \delta\Sigma_{\rm imp} 
\simeq \left[G^{0}_{\rm imp}\right]^2 \delta\Sigma_{\rm imp},
\end{align}
which coincides with the condition~\eqref{eq:SPcond} if the dual self-energy is taken in the form of  Eq.~\eqref{eq:Saddle-point-dual}.

\twocolumngrid

\bibliography{Ref}

\end{document}